\documentclass[english,aps,prd,superscriptaddress,preprintnumbers,nofootinbib,preprint]{revtex4-1}
\usepackage[T1]{fontenc}
\usepackage[utf8]{inputenc}
\usepackage{verbatim}

\usepackage{empheq}
\usepackage{amssymb}
\usepackage{amsmath}
\usepackage{amsfonts}

\usepackage[labelfont=bf,justification=raggedright]{caption}
\usepackage{subcaption}
\usepackage[colorlinks]{hyperref}
\hypersetup{
 citecolor=blue,
 linkcolor=blue,
 urlcolor=blue}

\usepackage{dcolumn}
\usepackage{graphicx}
\usepackage{etoolbox} 

\newcommand{\lp}{\left(}
\newcommand{\rp}{\right)}

\newcommand{\ep}{\epsilon}
\newcommand{\as}{\alpha_s}
\newcommand{\asb}{\alpha_{s,b}}
\newcommand{\asontwopi}{\left(\frac{\alpha_s}{2\pi}\right)}
\newcommand{\asbontwopi}{\left(\frac{\alpha_{s,b}S_{\ep}}{2\pi}\right)}
\newcommand{\nf}{n_f}
\newcommand{\tr}{T_R}
\newcommand{\Ca}{C_A}
\newcommand{\Cf}{C_F}
\newcommand{\MSbar}{\overline{\rm MS}}

\newcommand{\be}{\begin{equation}}
\newcommand{\ee}{\end{equation}}
\newcommand{\bea}{\begin{eqnarray}}
\newcommand{\eea}{\end{eqnarray}}
\newcommand{\bes}{\begin{equation}\begin{split}}

\newcommand{\LO}{\rm LO}
\newcommand{\NLO}{\rm NLO}
\newcommand{\NNLO}{\rm NNLO}

\newcommand{\fb}{\rm fb}

\newcommand{\gev}{\rm GeV}

\newcommand{\bb}{ {b\bar{b}} }

\newcommand{\BrHbb}{{\rm Br}(H\to\bb)}
\newcommand{\GamH  }{\Gamma_H}
\newcommand{\Gambb  }[1]{\ifblank{#1}{\Gamma_{\bb}}{\Gamma_{\bb}^{#1}}}
\newcommand{\dGambb  }[1]{\ifblank{#1}{{\rm d}\Gamma_{\bb}}{{\rm d}\Gamma_{\bb}^{#1}}}
\newcommand{\dgam    }[1]{\ifblank{#1}{{\rm d}\gamma}{{\rm d}\gamma^{#1}}}
\newcommand{\dsig    }[1]{\ifblank{#1}{{\rm d}\sigma}{{\rm d}\sigma^{#1}}}
\newcommand{\dsigWH  }[1]{\ifblank{#1}{{\rm d}\sigma_{WH}}{{\rm d}\sigma_{WH}^{#1}}}
\newcommand{\dsigWHbb}[1]{\ifblank{#1}{{\rm d}\sigma_{WH(\bb)}}{{\rm d}\sigma_{WH(\bb)}^{#1}}}

\newcommand{\kt}{k_t}
\newcommand{\mbb}{M_{H(\bb)}}
\newcommand{\dRbb}{\Delta R_{H(\bb)}}
\newcommand{\ptw}{p_{t,W}}
\newcommand{\pth}{p_{t,H(b\bar b)}}
\newcommand{\ptb}{p_{t,b}}
\newcommand{\ptWcut}{\ptw > 150~\gev}


\usepackage{babel}
\usepackage{color}
\begin{document}
\global\long\def\order#1{\mathcal{O}\left(#1\right)}
\global\long\def\d{\mathrm{d}}
\global\long\def\P{P}
\global\long\def\amp{{\mathcal M}}
\preprint{TTP20-011}
\preprint{P3H-20-009}
\preprint{CERN-TH-2020-043}

\def\KIT{
  Institute for Theoretical Particle Physics, KIT, Karlsruhe, Germany
}
\def\IKP{
  Institute for Nuclear Physics, KIT, Karlsruhe, Germany
}
\def\CERN{
  Theoretical Physics Department, CERN, 1211 Geneva 23, Switzerland
}
\def\OX{
  Rudolf Peierls Centre for Theoretical Physics, Clarendon Laboratory, Parks Road,
  Oxford OX1 3PU, UK and  Wadham College, Oxford OX1 3PN, UK
}

\title{
  Bottom quark mass effects in associated $WH$ production \\
  with $H \to \bb$ decay through NNLO QCD
}

\author{Arnd~Behring}
\email[Electronic address: ]{arnd.behring@kit.edu}
\affiliation{\KIT}

\author{Wojciech~Bizo\'{n}}
\email[Electronic address: ]{wojciech.bizon@kit.edu}
\affiliation{\KIT}
\affiliation{\IKP}

\author{Fabrizio~Caola}
\email[Electronic address: ]{fabrizio.caola@physics.ox.ac.uk}
\affiliation{\OX}

\author{Kirill~Melnikov}
\email[Electronic address: ]{kirill.melnikov@kit.edu}
\affiliation{\KIT}

\author{Raoul~R\"ontsch }
\email[Electronic address: ]{raoul.rontsch@cern.ch}
\affiliation{\CERN}

\begin{abstract}
  We present a computation of NNLO QCD corrections to the production
  of a Higgs boson in association with a $W$ boson at the LHC
  followed by the decay of the Higgs boson to a $\bb$ pair. At
  variance with previous NNLO QCD studies of the same process, we
  treat $b$ quarks as massive. An important advantage of working with
  massive $b$ quarks is that it makes the use of flavor jet algorithms
  unnecessary and allows us to employ conventional jet algorithms to
  define $b$ jets.  We compare NNLO QCD descriptions of the associated
  $WH(\bb)$ production with massive and massless $b$ quarks and also contrast
  them with  the results provided by parton showers.  We find ${\cal O}(5\%)$
  differences in fiducial cross sections computed with massless and massive $b$ quarks.
  We  also observe that much larger  differences
  between massless and massive results, as well as between fixed-order and parton-shower results,
  can arise in selected kinematic distributions.
\end{abstract}

\maketitle

\section{Introduction}

Detailed investigation of Higgs boson production in association with
a $W$ boson is an important part of the LHC research program that
aims at a comprehensive exploration of Higgs boson properties and
electroweak symmetry breaking
\cite{Chatrchyan:2013zna,Aaboud:2017xsd,Sirunyan:2017elk,Aaboud:2018zhk,Sirunyan:2018kst}. Indeed,
associated Higgs boson production gives us direct access to the $HWW$
coupling which is completely fixed in the Standard Model but can be
modified in its extensions. Moreover, studies of the $pp \to WH$ process
provide a unique way to study the Higgs coupling to $b$ quarks since, by
selecting Higgs bosons with relatively high transverse momenta, one
can identify $H\to\bb$ decays using substructure
techniques~\cite{Butterworth:2008iy,Marzani:2019hun}.

Interest in associated $WH$ production has inspired a large number of
theoretical computations that provide refined descriptions of this
process including
QCD~\cite{Han:1991ia,Baer:1992vx,Ohnemus:1992bd,Mrenna:1997wp,Spira:1997dg,Djouadi:1999ht,Brein:2003wg,Brein:2011vx,Brein:2012ne,Ferrera:2011bk,
  Ferrera:2013yga,Campbell:2016jau,Ferrera:2017zex,Caola:2017xuq,Gauld:2019yng} and
electroweak radiative
corrections~\cite{Ciccolini:2003jy,Denner:2011id}.  The more recent
theoretical efforts \cite{Ferrera:2011bk,
  Ferrera:2013yga,Campbell:2016jau,Ferrera:2017zex,Caola:2017xuq,Gauld:2019yng}
focused on a comprehensive fully-differential description of
associated production which consistently combines QCD
corrections to the production and decay processes.

All fully-differential NNLO QCD computations mentioned above have the
common feature that the decay of the Higgs boson to $b$ quarks is
described under the assumption that $b$ quarks are {\it massless}.
The same approximation is employed in the production subprocesses which
involve gluon splitting into a $\bb$ pair or $b$ quarks that come
directly from the proton.  Although, given the high energy of the LHC,
the massless approximation should be fairly adequate, there are a few
reasons that make it interesting to explore $b$-quark mass effects in
this process.

The first reason is that the phase space of the $pp \to WH$ process is large and
that there are important kinematic distributions which can be sensitive to
energy scales smaller than the total (partonic) energy of the
process. In those cases the dependence on the $b$-quark mass can
become more pronounced.  A comparison of fixed-order  computations,
including higher-order ones, for $pp \to WH$ performed with massless
and massive $b$ quarks will allow us to identify distributions and
phase-space regions with enhanced sensitivity to the $b$-quark mass.

The second reason to employ massive $b$ quarks in a calculation is
that in this case the splitting $g^* \to \bb $ becomes non-singular.
This feature makes it possible to employ {\it conventional} jet
algorithms to define $b$ jets.  We remind the reader that in case of
{\it massless} $b$ quarks, this is not possible and that a special
partonic flavor jet algorithm \cite{Banfi:2006hf} has  to be used.
The possibility to apply  conventional jet algorithms is an important
improvement since  it makes theoretical computations and experimental
analyses more aligned.

The third reason is the appearance of certain contributions in the $H
\to \bb$ decay which cannot be properly described if $b$ quarks are
treated as massless.  It was pointed out in Ref.~\cite{Caola:2017xuq}
that an interference of singlet $H \to g(g^* \to \bb)$ and non-singlet
$H \to (b^* \to bg) \bar b$ decay amplitudes forces us to keep the
mass of the $b$ quark different from zero throughout the computation
since otherwise unconventional soft quark divergences appear. Such
studies have already been carried out in Ref.~\cite{Primo:2018zby}.

Motivated by these considerations, we extended the computation of NNLO
QCD radiative corrections reported in Ref.~\cite{Caola:2017xuq} to
include $b$-quark mass effects in the theoretical description of Higgs
production in association with a vector boson. To this end, we
combined the recent NNLO QCD description of the Higgs boson decay into
two massive $b$ quarks~\cite{Behring:2019oci}\footnote{A 
  calculation of NNLO QCD corrections to the $H\to\bb$ decay with
  massive $b$ quarks was also performed in
  Ref.~\cite{Bernreuther:2018ynm}.} with the computation of NNLO QCD
corrections to the production process
~\cite{Caola:2017xuq,Caola:2019nzf} which required small modifications
because of the $b$-quark mass.

In addition to fixed-order computations, parton showers are widely
used to provide theoretical predictions for collider experiments.
In the context of associated Higgs production, they have been employed
in
Refs.~\cite{Frixione:2005gz,Hamilton:2009za,Granata:2017iod,Luisoni:2013kna,Astill:2018ivh,Bizon:2019tfo,Alioli:2019qzz}.
For this reason, it is interesting to compare fixed-order and
parton-shower results with each other.
Although this has already been done in Ref.~\cite{Caola:2017xuq}, the
need to use different jet algorithms in fixed-order massless and
parton-shower computations did not allow a direct comparison of the
two.  The NNLO QCD computation with massive $b$ quarks
described in this paper allows us to remedy this problem and compare
fixed-order and parton-shower predictions using identical jet
algorithms.

The remainder of the paper is organized as follows.  In
Section~\ref{sect:basics} we briefly review the NNLO QCD computation
of radiative corrections to $pp \to WH$~\cite{Caola:2017xuq} and
$H \to \bb$~\cite{Behring:2019oci} and discuss modifications needed in
the computation of NNLO QCD corrections to the production process to
accommodate massive $b$ quarks.  In Section~\ref{sect:WHprod} we show
numerical results for NNLO QCD corrections to $pp \to WH(\bb)$ with
massive $b$ quarks and compare them to results of the massless
computation.  In Section~\ref{sect:shower} we compare a parton-shower description of associated $WH$ production
with fixed-order results.   We conclude in Section~\ref{sect:concl}. A detailed discussion of modifications required to accommodate
massive $b$ quarks in the NNLO QCD computation of Ref.~\cite{Caola:2017xuq} can be found in two  appendices.

\section{Summary of NNLO QCD computations}
\label{sect:basics}

In this section, we briefly review the computation of NNLO QCD
radiative corrections to the associated production $pp \to WH$ and the
$H \to \bb$ decay processes.
An earlier computation of NNLO QCD corrections to $pp \to WH$
 was
described in Ref.~\cite{Caola:2017xuq} using the formulation of the nested soft-collinear subtraction scheme presented in Ref.~\cite{Caola:2017dug}. Since then, simple analytic
formulas for the NNLO QCD corrections to the production of a color-singlet final state in hadron
collisions were published in Ref.~\cite{Caola:2019nzf}. These formulas
employ results for integrated double-unresolved soft and collinear
subtraction terms computed in Refs.~\cite{Caola:2018pxp} and
\cite{Delto:2019asp}, respectively.  To accommodate these
developments, the code that allows us to compute NNLO QCD corrections
to $pp \to WH$ was updated. In addition, we refined the description of
the $H \to \bb$ decay with massless $b$ quarks using analytic
results for NNLO QCD corrections to decays of color-singlet particles
derived in Ref.~\cite{Caola:2019pfz}.

To accommodate massive $b$ quarks, we employed a recent
computation~\cite{Behring:2019oci} of the NNLO QCD corrections to
$H \to \bb$ that fully accounts for the $b$-quark mass. That
computation is based on the nested soft-collinear subtraction scheme
adapted to deal with massive particles.  On the production side,
a consistent description of $b$ quarks as massive particles forces us to
work in a four-flavor scheme so that $b$ quarks are excluded from
parton distributions.  This feature leads to some changes to the
renormalization procedure that we discuss in
Appendix~\ref{app:renorm}.  In addition, we have to modify the
computation of NNLO QCD corrections to $pp \to WH$ to describe the
splitting of a gluon into a massive $\bb$ pair, and the gluon vacuum
polarization contributions due to  massive $b$-quark loop.

We note that the gluon splitting contribution refers to the process $q_i q_j \to WH + ( g^* \to \bb)$
which is free of soft and  collinear singularities thanks to
the finite mass of the $b$ quark. The resulting logarithmically enhanced
terms of the form $\log (s/m_b^2)$ may, potentially, be large, but they
do not appear to be particularly problematic from a numerical
viewpoint. Hence, to describe these contributions, we calculate
helicity amplitudes for the $q_i q_j \to WH \bb$ process, parametrize the
$WH \bb$ phase space and perform numerical integration to compute
observables of our choice.

Two-loop corrections to the $q_i \bar q_j W$ vertex caused by the gluon
vacuum polarization due to a massive quark loop
 can be extracted from
Refs.~\cite{Kniehl:1989kz,Rijken:1995gi,Blumlein:2016xcy}.  We recomputed these contributions and
found full agreement with the results presented in
Ref.~\cite{Kniehl:1989kz}.  For completeness, we provide the details
of our calculation in~Appendix~\ref{app:formf}.

\section{ The process $pp \to WH(\bb)$ }
\label{sect:WHprod}

In this section we present results for the associated production $pp \to WH(\bb)$
including $b$-quark mass effects. We begin by specifying how corrections
to production and decay processes are combined.
Since the Higgs boson is a scalar particle, these
corrections can be put together in a straightforward manner. The only
subtlety worth discussing is how to treat the {\it total} decay width
of the Higgs boson that appears in the differential cross section for
$pp \to WH(\bb)$ when it is computed in the narrow-width
approximation.
We begin by writing the cross section as follows~\cite{Ferrera:2013yga}
\begin{align}
  \dsigWHbb{}
  ={}&
       \dsigWH{}
       \times
       \frac{ \dGambb{} }{ \GamH{} }
  ={}
       \BrHbb{}
       \times
       \dsigWH{}
       \times
       \frac{ \dGambb{} }{ \Gambb{} }
       \,.
\label{eq:sigWHbb}
\end{align}
We treat $\BrHbb$ as an input parameter and do not expand it in a series
in $\alpha_s$\footnote{We note that other choices are possible, see Ref.~\cite{Gauld:2019yng} for a comprehensive discussion.}.  For numerical computations we take $\BrHbb = 0.5824$,
as recommended by the Higgs Cross Section Working
Group~\cite{deFlorian:2016spz}.

Keeping the branching fraction fixed, we compute an expansion of
Eq.~\eqref{eq:sigWHbb} in $\alpha_s$ by first expanding the $WH$ cross
section and the decay rate
\begin{align}
  \dsigWH{}
  ={}&
       \sum_{i=0}^{\infty}
       \dsigWH{(i)}\,, &
  \dGambb{}
  ={}&
       \sum_{i=0}^{\infty}
       \dGambb{(i)}\,,
\end{align}
then introducing {\it normalized} quantities to describe the decays
\begin{align}
  \dgam{(i)}
  ={}&
       \frac{
       \sum\limits_{k=0}^{i} \dGambb{(k)}
       }{
       \sum\limits_{k=0}^{i} \Gambb{(k)}
       }\,,
       \label{eq:gamNorm}
\end{align}
and, finally, defining physical
cross sections computed through different orders in QCD perturbation
theory
\begin{align}
  \begin{aligned}
    \dsigWHbb{\LO}
    ={}&
    \BrHbb{}
    \big[
    \dsig{(0)}\dgam{(0)}
    \big]
    \,,\\
    \dsigWHbb{\NLO}
    ={}&
    \BrHbb{}
    \big[
    \dsig{(0)}\dgam{(1)}
    +\dsig{(1)}\dgam{(0)}
    \big]
    \,,\\
    \dsigWHbb{\NNLO}
    ={}&
    \BrHbb{}
    \big[
     \dsig{(0)}\dgam{(2)}
    +\dsig{(1)}\dgam{(1)}
    +\dsig{(2)}\dgam{(0)}
    \big]
    \,.
    \label{eq:sigExp}
  \end{aligned}
\end{align}
We note that ${\int } \dgam{(i)} = 1$ provided that the integration is
performed over unrestricted phase space. An identical definition of the cross
section was used in an earlier massless computation reported in
Ref.~\cite{Caola:2017xuq}.

To present the results of our computation,  we focus on the associated production process
\begin{align}
  p p
  \to
  W^+ H
  \to
  (\nu_e \bar{e})
  (\bb ).
\end{align}
We treat both decay processes $W^+ \to \nu_e \bar e$ and $ H \to \bb$
in the narrow-width approximation.  We set the Higgs boson mass to
$M_H = 125~\gev$, the $W$-boson mass to $ M_W = 80.399~\gev$ and the {\it
  on-shell} $b$-quark mass to $m_b = 4.78~\gev$. We note that the $b$-quark
Yukawa coupling that enters the $H \to \bb$ decay is computed using
the $\MSbar$ $b$-quark mass calculated at $\mu = M_H$. However, since
physical cross sections in Eq.~\eqref{eq:sigExp} are proportional to
the ratio $\dGambb{}/\Gambb{}$, the dependence on the Yukawa coupling
cancels out (almost) completely\footnote{At NNLO a residual dependence
  on $y_b$ remains in the $\dGambb{}/\Gambb{}$
  ratio because of the singlet-non-singlet interference which depends
  on the product of top and bottom Yukawa couplings.}  in the results
that are presented below. The top-quark mass is set to
$m_t = 173.2~\gev$. We use the Fermi constant
$G_F = 1.16639 \times 10^{-5}~\gev^{-2}$ and the sine squared of the
weak mixing angle   $\sin^2\theta_W = 0.2226459$. The width of the $W$ boson is taken to be $\Gamma_W = 2.1054~\gev$. Finally, we
approximate the CKM matrix by an identity matrix.\footnote{We have checked through NLO QCD
  that in case of the associated
  production, this approximation is accurate to about a percent.}

We also need to specify the selection criteria that are used to define
the $W(\nu_e \bar e)\; H(\bb)$ final state.  To this end, we require
that an event contains at least two $b$ jets that are defined with the
anti-$\kt$ jet algorithm \cite{Cacciari:2008gp,Cacciari:2011ma}. For the sake of
comparison, we also calculate $WH(\bb)$ cross sections for massless $b$
quarks. In that case, we employ the {\it flavor}-$\kt$ jet algorithm
\cite{Banfi:2006hf} to describe massless $b$ jets. In both cases, we
choose the jet radius $R=0.4$.  Moreover, we impose the following cuts
on pseudo-rapidities and transverse momenta of leptons and $b$ jets
\begin{align}
  \begin{aligned}
    &|\eta_{l}|   < 2.5 \,, \quad p_{t,l} > 15~\gev\,,\\
    &|\eta_{j,b}| < 2.5 \,, \quad p_{t,jb} > 25~\gev\,.
  \end{aligned}
\end{align}
Finally, following experimental analyses, we may employ an additional
requirement that the vector boson has a transverse momentum of
$\ptWcut$. We always state explicitly when this
cut is used.

To present numerical results we use the five- and four-flavor parton
distribution function sets \verb+NNPDF31_nnlo_as_0118+ and
\verb+NNPDF31_nnlo_as_0118_nf_4+ for computations with
massless and massive $b$ quarks, respectively.  We employ NNLO PDFs
to compute LO, NLO and NNLO cross sections in what follows.
Moreover, both in massive and massless cases, we use $\as(M_Z) =
0.118$ and perform the running of the strong coupling at three loops
with five active flavors.\footnote{We note that, to be fully
  consistent, one should use doped parton distribution
  functions~\cite{Bertone:2015gba}. We defer this for future work.}

For all cross sections the central value of the renormalization and
factorization scales {\it in the production process} is set to one
half of the invariant mass of the $WH$ system, i.e.
$\mu_r = \mu_f = \tfrac{1}{2} \sqrt{(p_W+p_H)^2}$, whereas the
renormalization scale {\it for the decay process} is set to the Higgs
boson mass, $\mu_{r,\rm{dec}} = M_H$.
The uncertainty of the cross sections is obtained by varying the scale
in the production process by a factor of two and, independently,
by changing the decay scale by a factor of two as well. The total
uncertainty is taken to be an envelope of these nine numbers.

\begin{table}
  \centering
  \begin{tabular}{llcc}
    \hline
    Order
    & $b$ quarks
    & $\sigma_{\rm fid}~[\fb]$
    & $\sigma_{\rm fid}({\rm boosted})~[\fb]$
    \\
    \hline
    $\LO$
    & massive
    & $  22.623^{  +0.845}_{  -1.047}$
    & $   3.735^{  +0.000}_{  -0.016}$
    \\
    $ $
    & massless
    & $  22.501^{  +0.796}_{  -1.007}$
    & $   3.638^{  +0.000}_{  -0.009}$
    \\
    \hline
    $\NLO$
    & massive
    & $  25.364(1)^{  +0.778}_{  -0.756}$
    & $   4.586(1)^{  +0.158}_{  -0.141}$
    \\
    $ $
    & massless
    & $  24.421(1)^{  +0.852}_{  -0.879}$
    & $   4.333(1)^{  +0.165}_{  -0.154}$
    \\
    \hline
    $\NNLO$
    & massive
    & $  24.225(4)^{  +0.642}_{  -0.742}$
    & $   4.530(2)^{  +0.071}_{  -0.096}$
    \\
    $ $
    & massless
    & $  22.781(3)^{  +0.791}_{  -0.898}$
    & $   4.207(1)^{  +0.097}_{  -0.116}$
    \\
    \hline
  \end{tabular}
  \caption{Fiducial cross sections for $pp \to W^+H(\bb)$ at the $13~{\rm TeV}$ LHC at various orders of  QCD
    perturbation
    theory calculated for massive and massless $b$ quarks.  The label
    ``boosted'' implies that an  additional cut is imposed on the $W$ boson's
    transverse momentum, $\ptWcut$.  The uncertainty is estimated
    using scale variation. The numerical integration error is reported in round
    brackets. See main text for details.}
  \label{tab:fid-xsec}
\end{table}
We begin by presenting fiducial cross sections for the process
$pp \to W^+ H(\bb)$ at the $13$~TeV LHC  in Table~\ref{tab:fid-xsec}.
Comparing these results with massless predictions, we observe that the
massive cross sections are systematically larger than the massless
ones.
The difference is very small at LO but increases when radiative
corrections are included.  At NLO, the differences range from about four
percent, in case of the basic fiducial cuts, to six percent, if the
additional $\ptWcut$ cut is applied.  At NNLO, the differences between
massive and massless results increase further and reach $6-7$ percent.

We note that these differences may be obscured by the scale variation
uncertainties.  This is indeed what happens at leading and, to some
extent, also at next-to-leading order, whereas at NNLO the massive and
massless cross sections differ from each other even if their scale
variation uncertainties are accounted for.
  
We emphasize that the NNLO scale variation uncertainties shown in
Table~\ref{tab:fid-xsec} are likely to be too
conservative~\cite{Gauld:2019yng}.  Indeed, it was shown in
Ref.~\cite{Gauld:2019yng} that upon including a perturbative expansion
of the $H \to b \bar b$ branching ratio in the definition of ${\rm d}
\sigma_{WH}$, see Eq.~\eqref{eq:gamNorm}, the NNLO scale uncertainty
of the so-defined cross section reduces to a sub-percent level and
becomes close to the uncertainty that is associated with the scale
variation in the $WH$ production process without the decay. With this
in mind, when discussing kinematic distributions, we only show results
obtained with the central scale choice.

The ${\cal O}(5\%)$ differences between massive and massless
fiducial cross sections can be traced back to gluon radiation in
$H\to\bb$ decays. Indeed, it is well-known that the collinear
radiation pattern of massive and massless $b$ quarks differs
significantly. In case of massless $b$ quarks, we expect a logarithmic
enhancement of the collinear gluon emission probability ${\rm d}{\cal
  P} \sim {\rm d} \theta^2/\theta^2$, where $\theta$ is the relative
angle between the $b$ quark and the gluon momenta.  This feature leads to
a logarithmic dependence of the fiducial cross section on the
clustering radius $R$. At the same time, when massive $b$ quarks
radiate, the probability distribution becomes $ {\rm d}{\cal P} \sim
{\rm d} \theta^2/( \theta^2 + m_b^2/E_b^2)$, where $E_b$ is the energy
of the radiating quark. This probability distribution implies that the
collinear singularity is screened by the $b$-quark mass and that the
cross-section dependence on the jet radius changes if $\Delta R <
m_b/\ptb$.
We have checked that for the chosen value of the jet radius, the
amount of radiation included in a $b$ jet is
different for massless and massive quarks. This means that, in the case of
radiative decay of the Higgs boson $H \to \bb g$, the acceptance of
events with massless $b$ quarks is smaller, by about a factor of two,
than the acceptance computed with massive $b$ quarks, when fiducial
cuts described above are applied.
We also observe that this difference is reduced if we consider larger
jet radii or reduce the transverse momentum cut on $b$ jets.

Finally, it turns out that the ${\cal O}(y_t y_b)$ interference of
singlet $H \to g(g^* \to \bb)$ and non-singlet $H \to (b^* \to bg) \bar
b$ decay amplitudes, discussed in Ref.~\cite{Caola:2017xuq}, is a
minor effect.  For the fiducial cuts discussed above, it contributes
to cross sections only at a sub-percent level and is, therefore, below
the scale uncertainty and much smaller than the differences between
massive and massless computations.

We will now proceed with the discussion of kinematic
distributions. Since in an experimental analysis the Higgs boson can
only be observed through its decay products, we will study kinematic
distributions of the $b$- and $\bar b$-jet pair whose invariant mass is
closest to the Higgs boson mass.  Throughout this paper, we refer to
such pairs of jets with the subscript $H(\bb)$, e.g. their four-momenta
are written as $p_{H(\bb)}$ and their invariant masses as $M_{H(\bb)}$.

\begin{figure}\centering
  \includegraphics[width=0.45\textwidth]{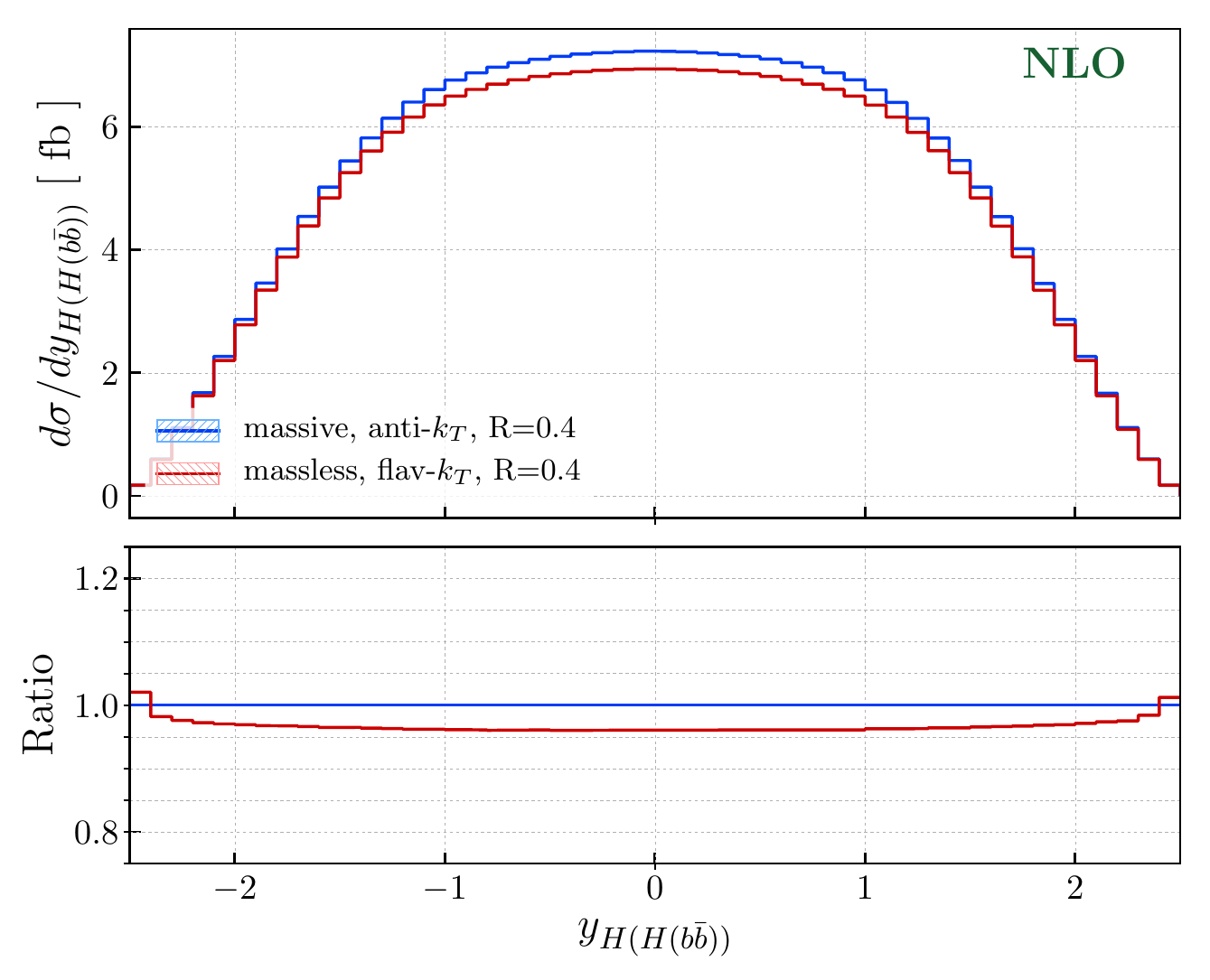}
  \includegraphics[width=0.45\textwidth]{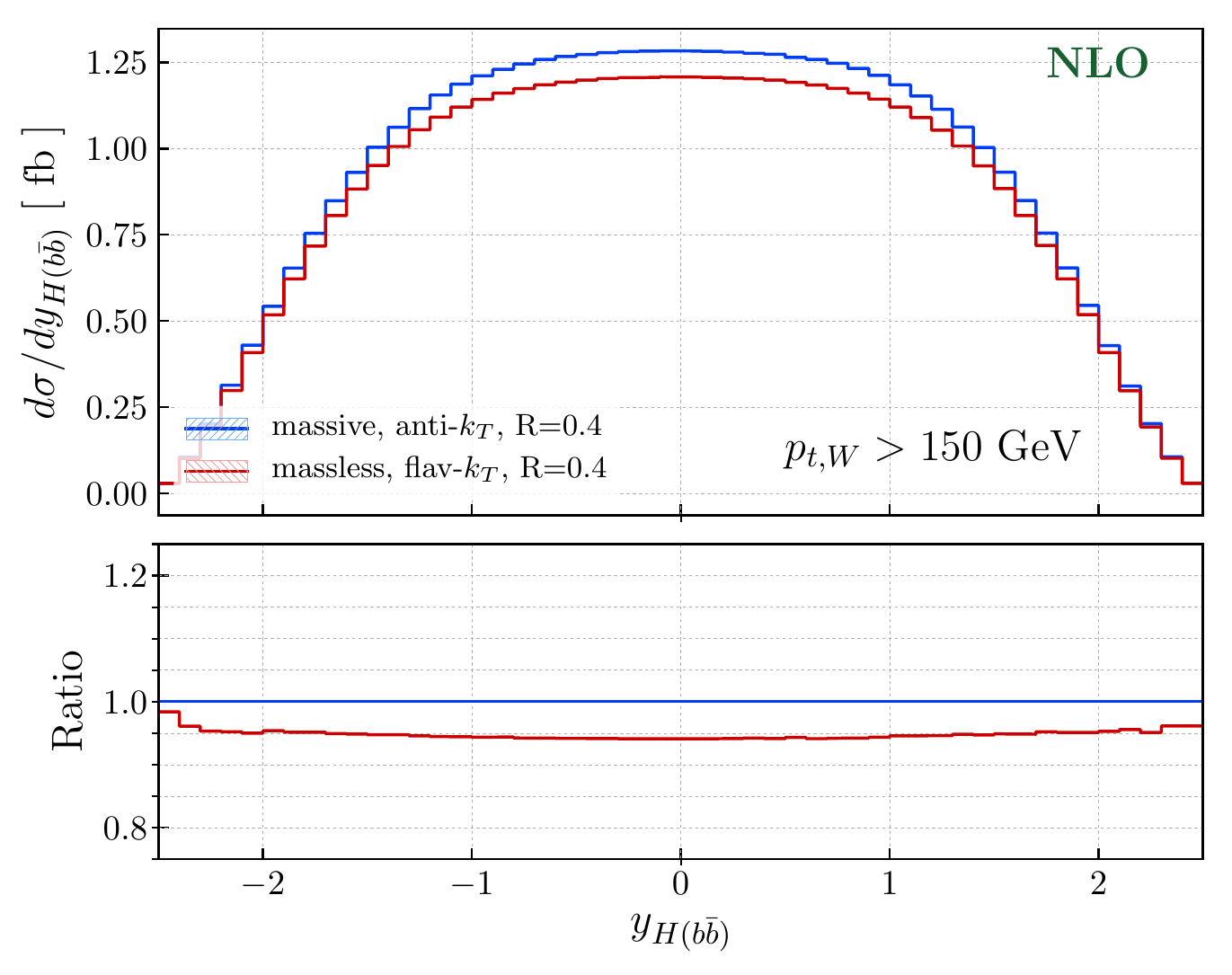}
  \includegraphics[width=0.45\textwidth]{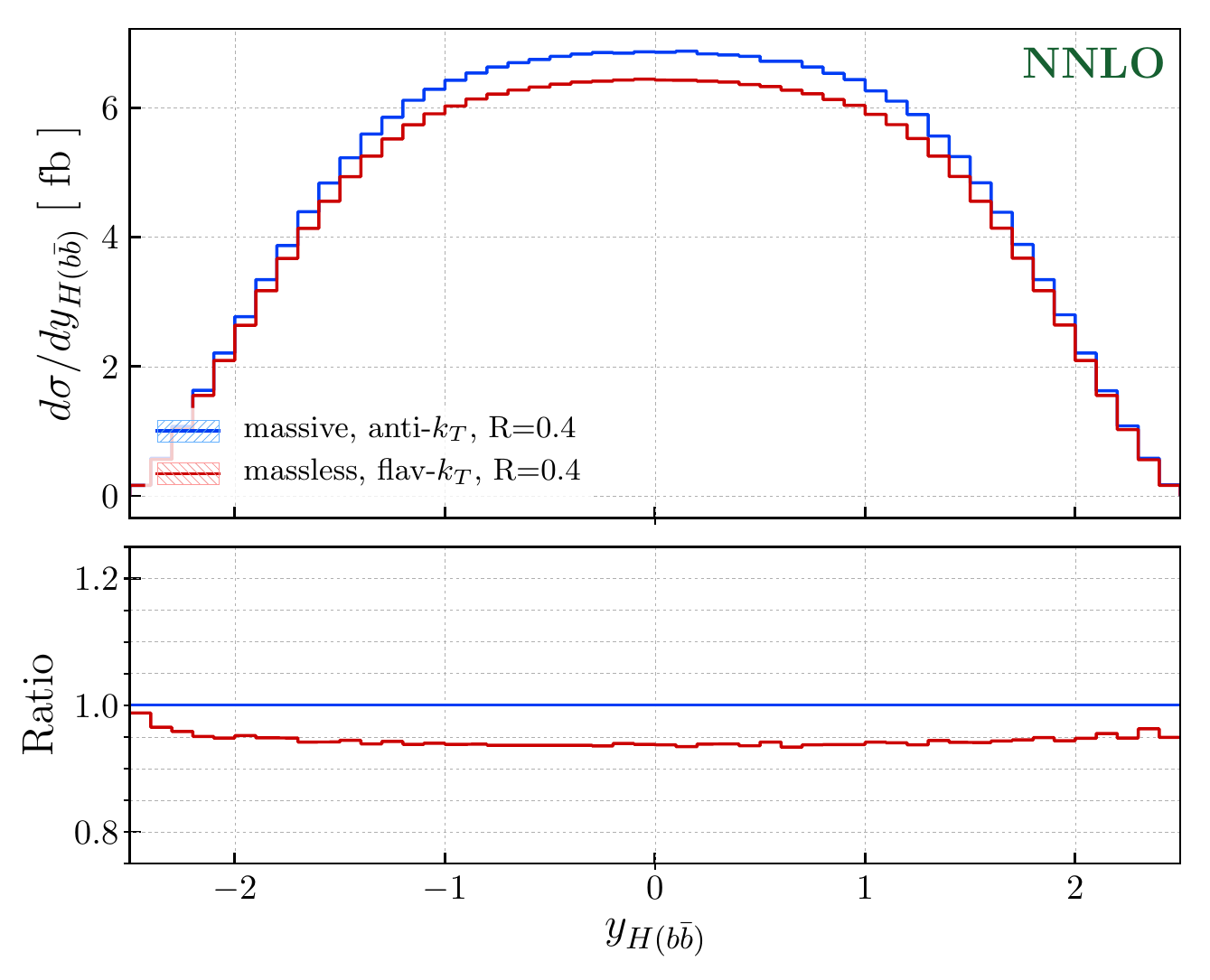}
  \includegraphics[width=0.45\textwidth]{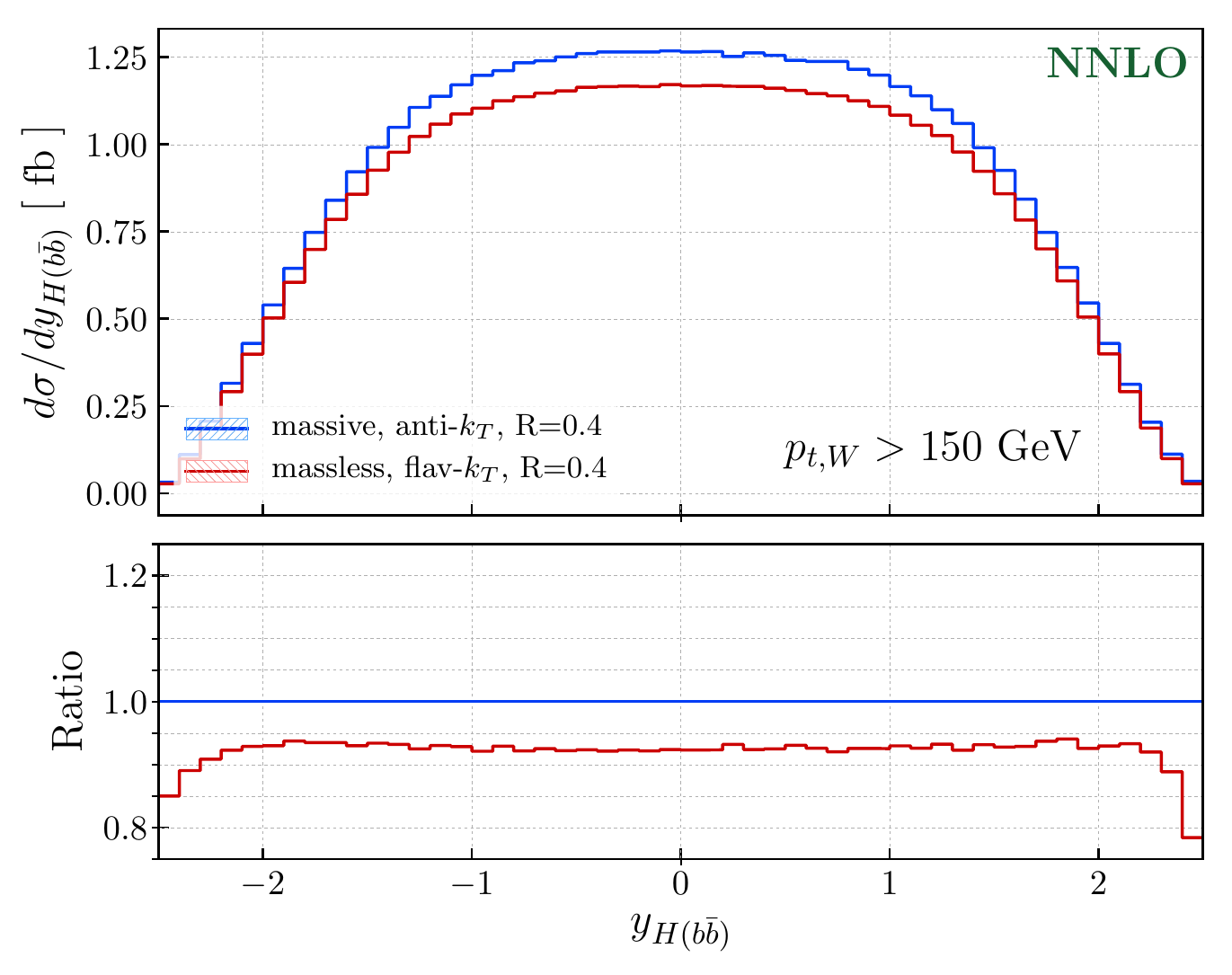}
  \caption{The rapidity distribution of the reconstructed Higgs boson
    calculated at NLO (upper plots) and NNLO (lower plots) for central
    values of the renormalization and factorization scales. Lower
    panes show ratios of massless to massive results.  See text for
    details.  }
  \label{fig:Hrap}
\end{figure}

We begin  by presenting the rapidity distribution of pairs of $b$ jets
in Fig.~\ref{fig:Hrap}.
We observe that the distributions computed with massive and massless $b$
quarks are very similar and differ, to a good approximation, by an
overall rescaling factor that can be inferred from the results for the cross
sections reported in Table~\ref{tab:fid-xsec}.  Such behavior is
expected given the well-known inclusiveness of rapidity distributions.

\begin{figure}
  \centering
  \includegraphics[width=0.45\textwidth]{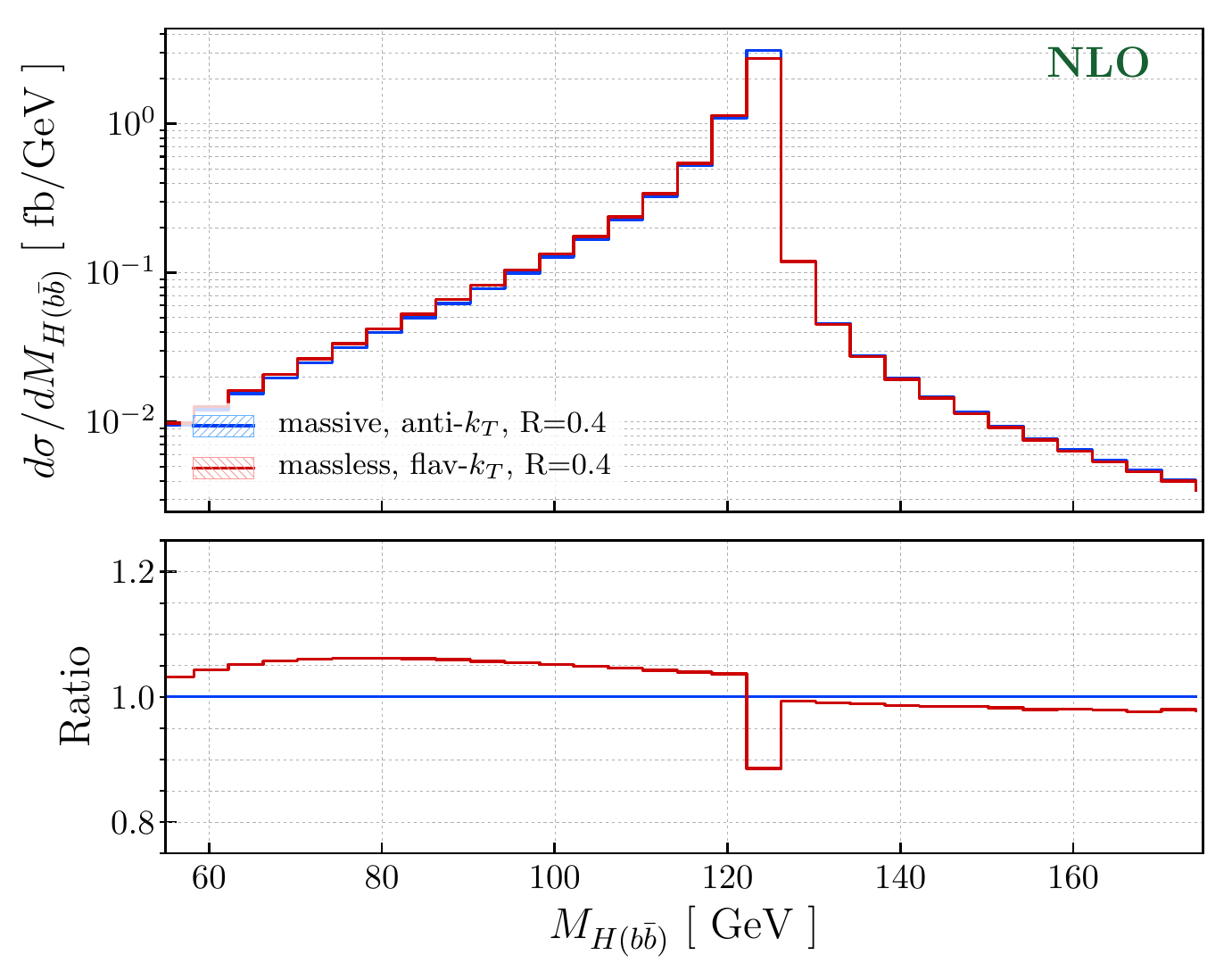}
  \includegraphics[width=0.45\textwidth]{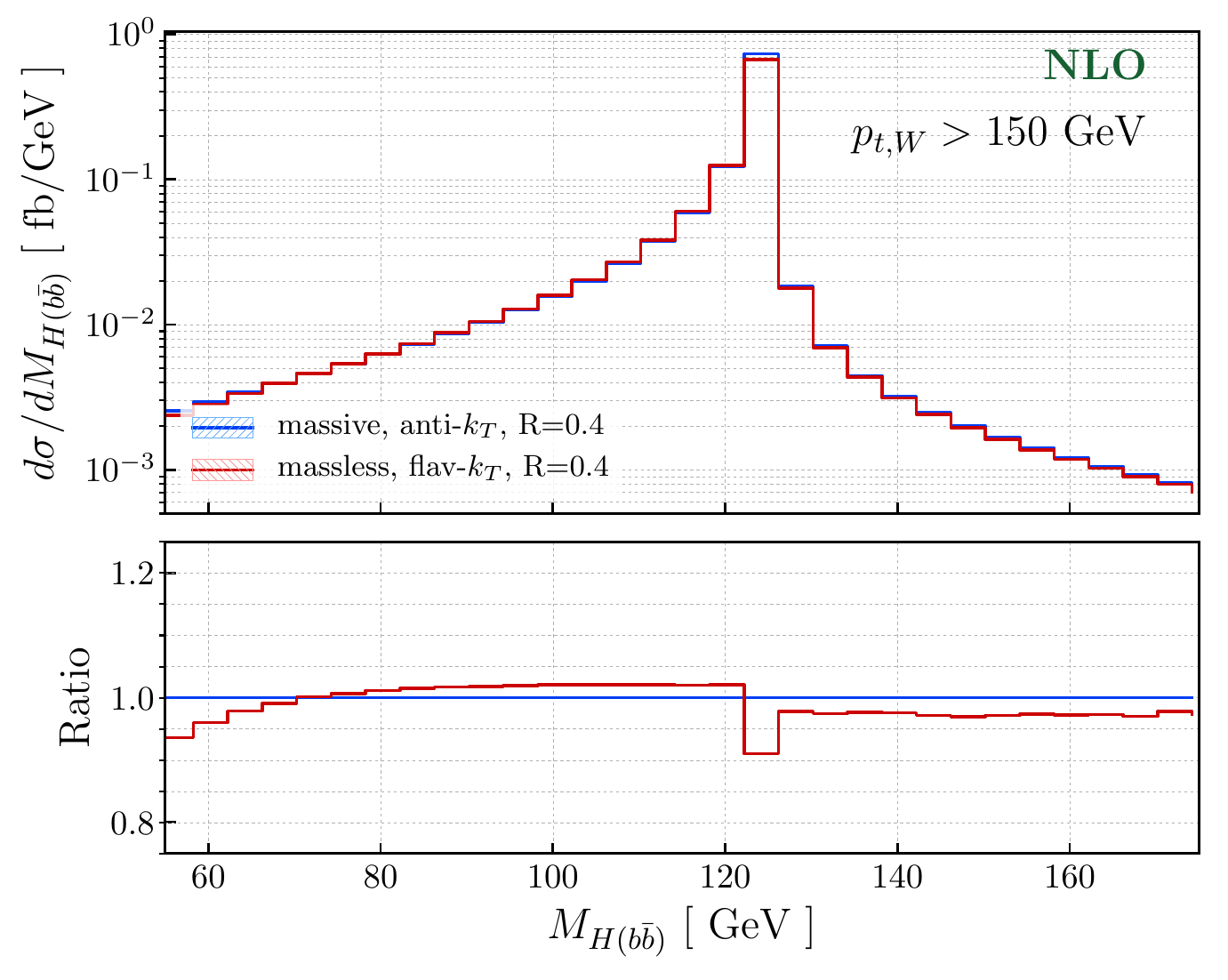}
  \includegraphics[width=0.45\textwidth]{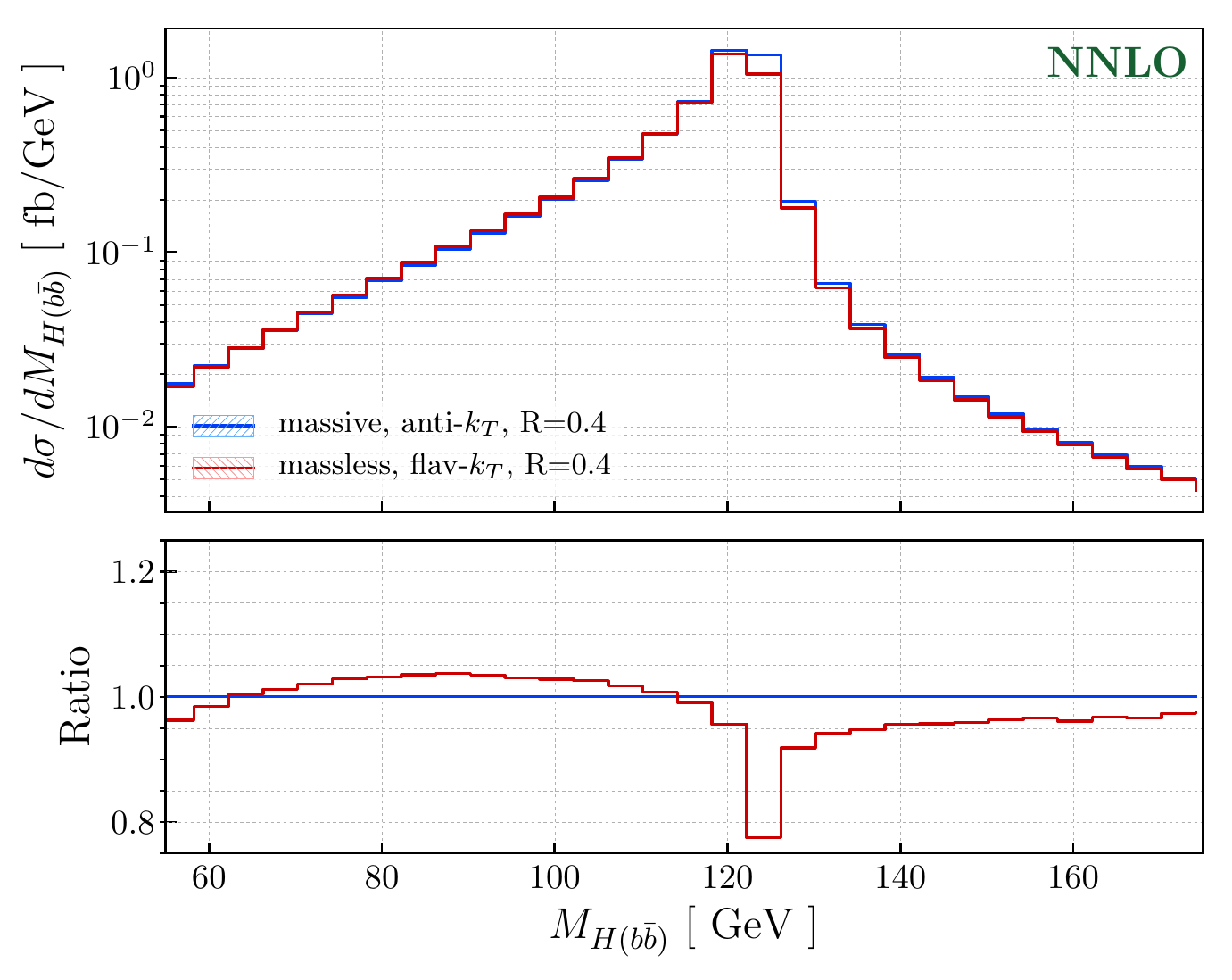}
  \includegraphics[width=0.45\textwidth]{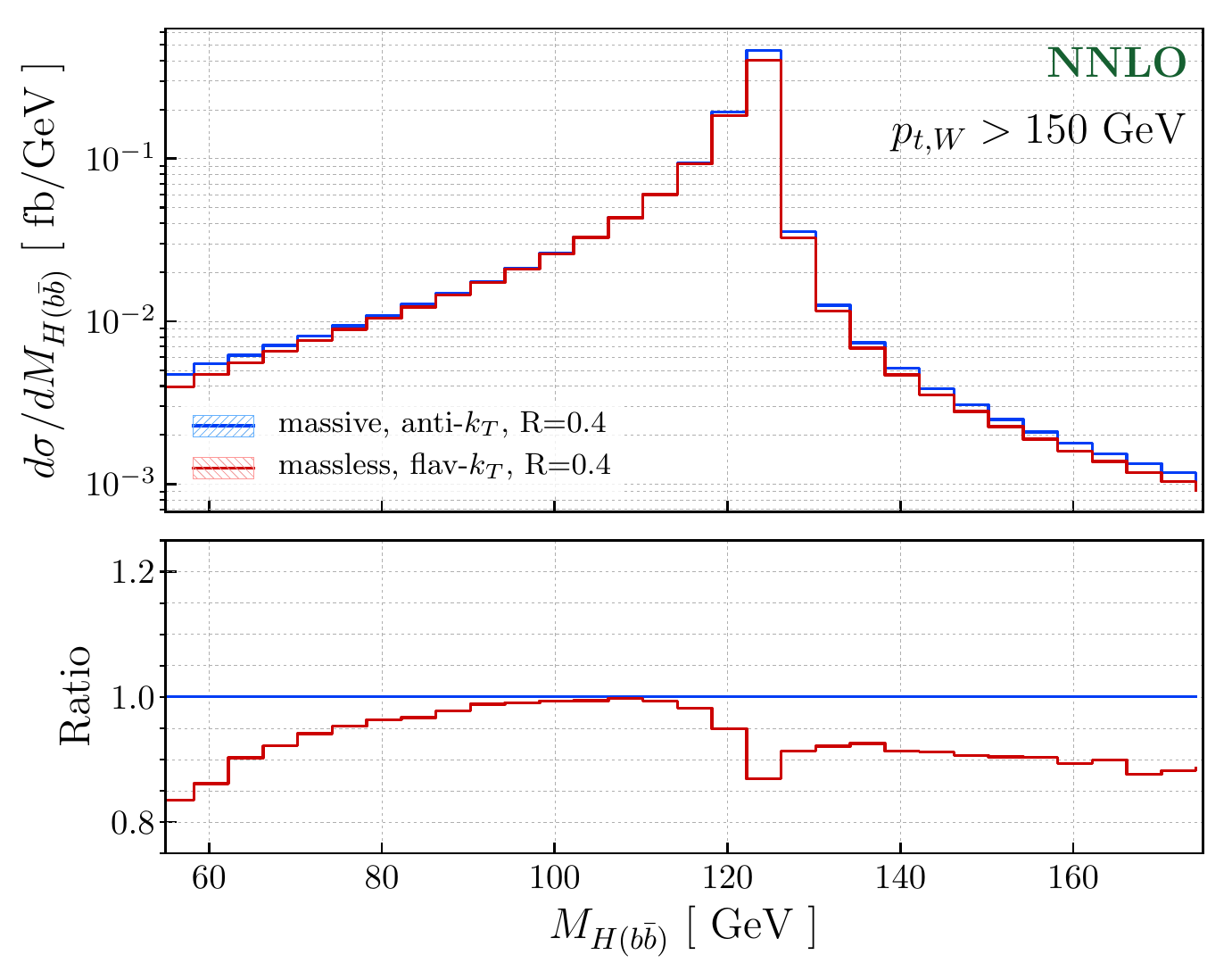}
  \caption{The invariant mass distribution of the two $b$ jets that
    best reconstruct the Higgs boson mass calculated at NLO (upper
    plots) and NNLO (lower plots) for central values of the renormalization and factorization scales.
Lower panes
    show ratios of massless to massive results.
 See text for details.
  }
  \label{fig:mHbb}
\end{figure}
We proceed with the invariant mass distribution of the two $b$ jets, $M_{H(\bb)}$, which is presented
in  Fig.~\ref{fig:mHbb}.
At leading order this distribution is described by a $\delta$-function,
$\delta( \mbb^2 - M_H^2)$, but the situation becomes more complex when
higher-order corrections are considered.
In particular, if a $b$ quark from the decay is clustered with a gluon
emitted in the production process, the invariant mass of two $b$ jets
can exceed $M_H$ and, conversely, a three-body decay
$H \to \bb g$ leads to two $b$ jets with an invariant mass that
is  smaller than $M_H$. Hence, already at NLO, the $\mbb$
distribution is non-vanishing both below and above $M_H$.
We present the $\mbb$ distributions obtained at NLO and NNLO in
Fig.~\ref{fig:mHbb}.
If the $\ptWcut$ cut is not applied, we observe that below the Higgs
peak, the massless results are larger than the massive ones except at
very low invariant masses. In the region above the peak, which is
populated by events with radiative corrections to the production
process, the two results are very similar to each other. In the most
populated bin, adjacent to the Higgs boson mass, $\mbb=M_H$, the
massive result is larger than the massless one; this feature drives
the observed behavior for fiducial cross sections discussed earlier,
c.f.  Table~\ref{tab:fid-xsec}.
When the additional $\ptWcut$ cut is applied, the massless result
stays below the massive one; we observe an $\mathcal{O}(15\%)$
difference at very low invariant masses which decreases when getting
closer to the peak. Above the Higgs mass, we see a constant
difference of about $10\%$.

\begin{figure}\centering
  \includegraphics[width=0.45\textwidth]{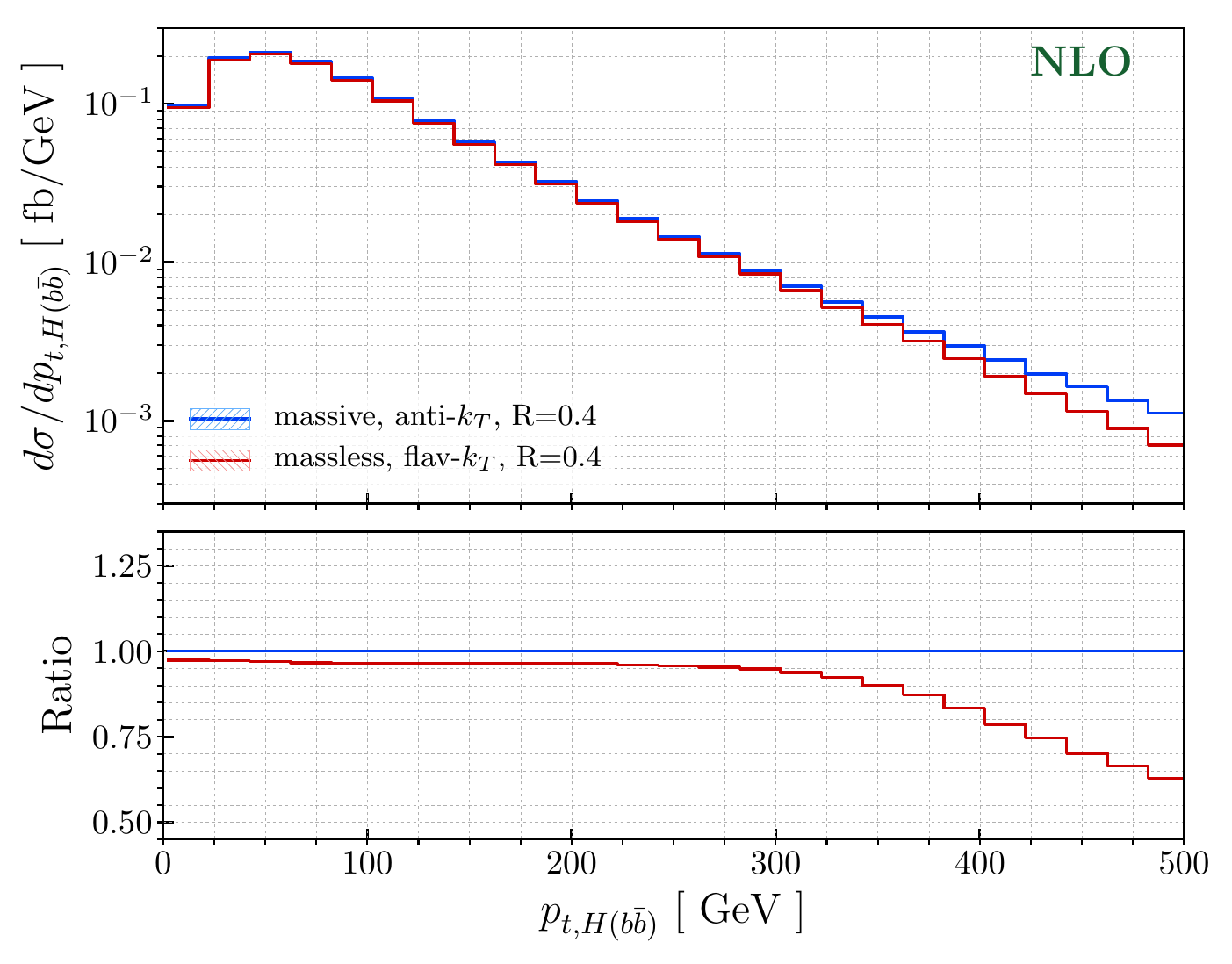}
  \includegraphics[width=0.45\textwidth]{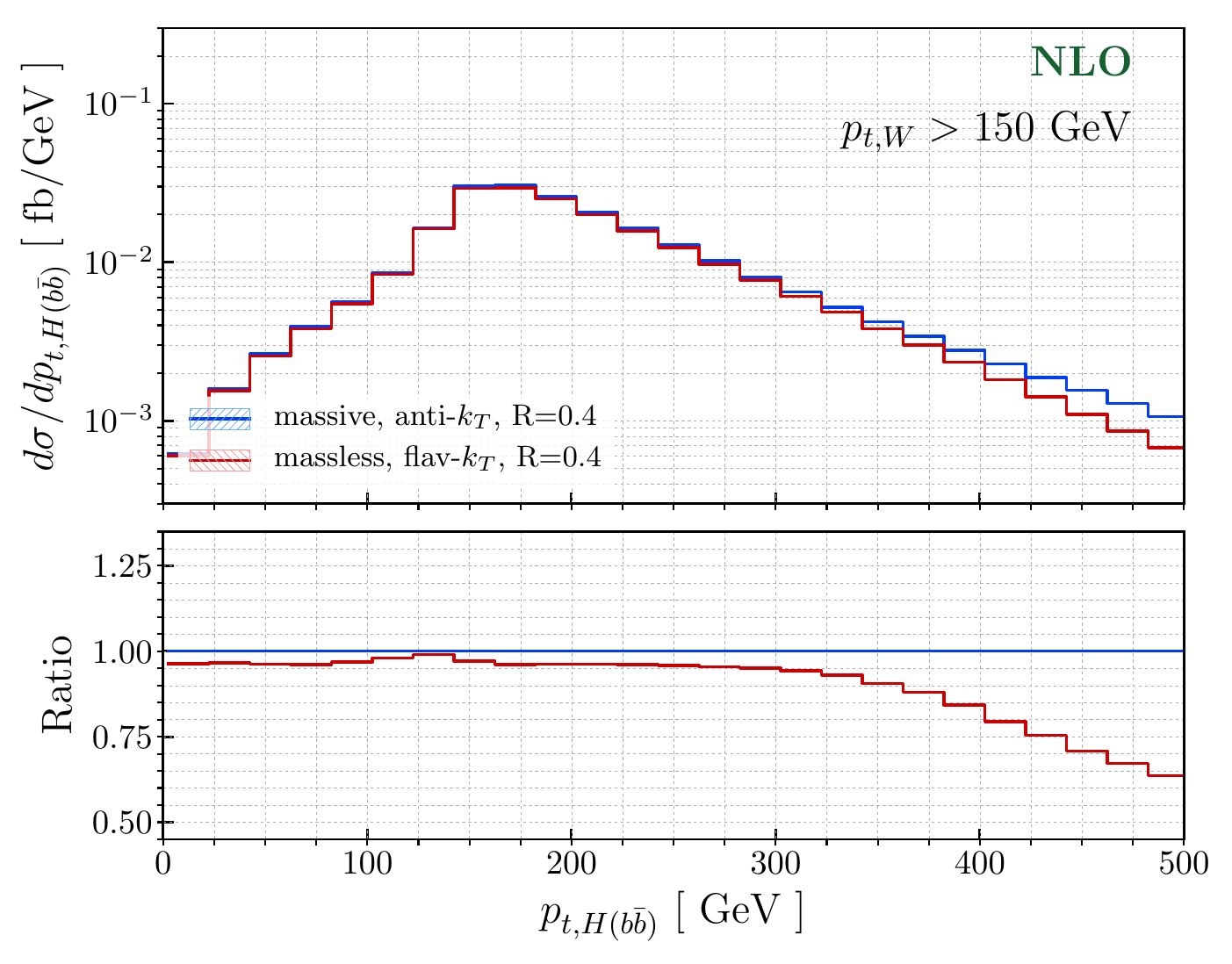}
  \includegraphics[width=0.45\textwidth]{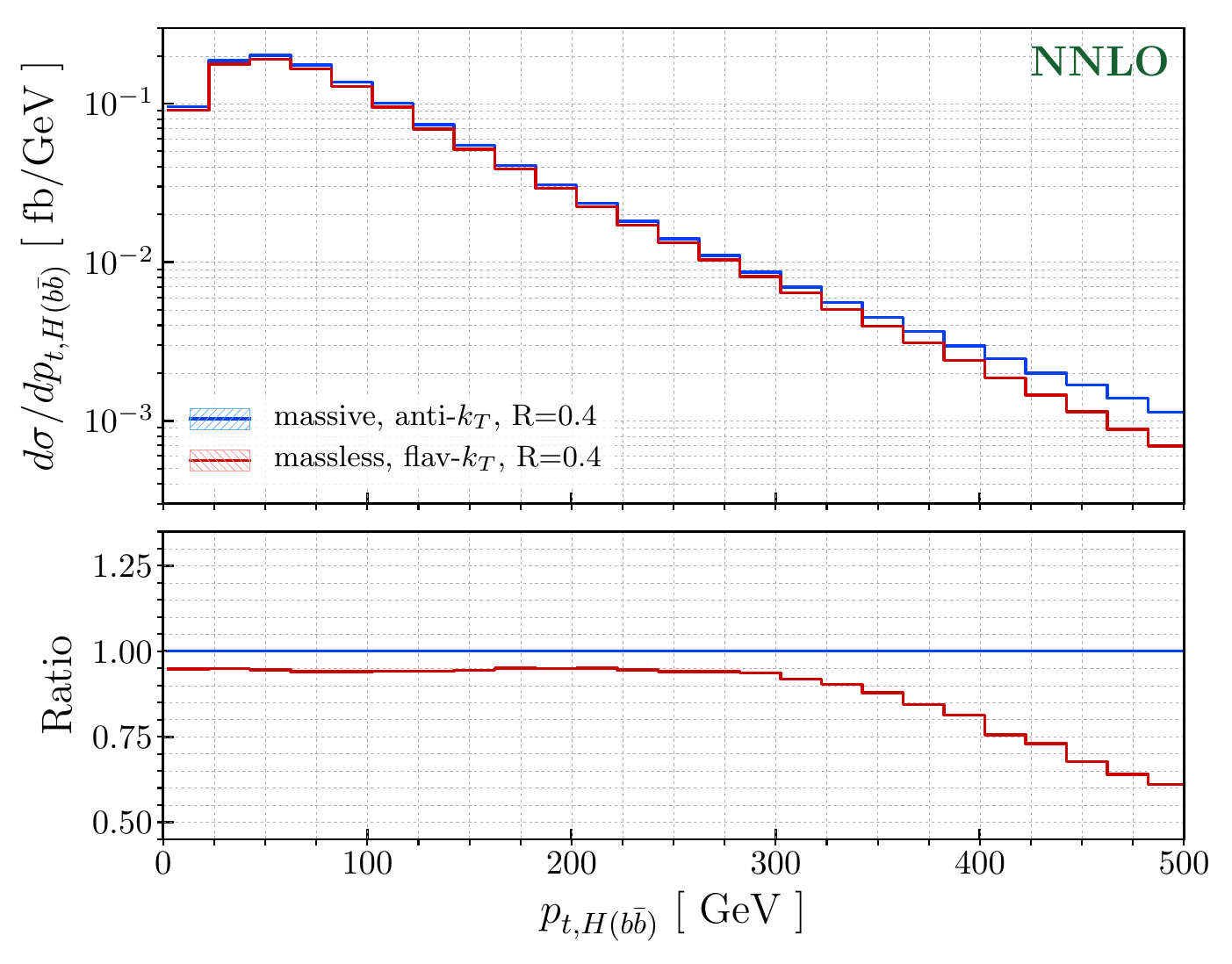}
  \includegraphics[width=0.45\textwidth]{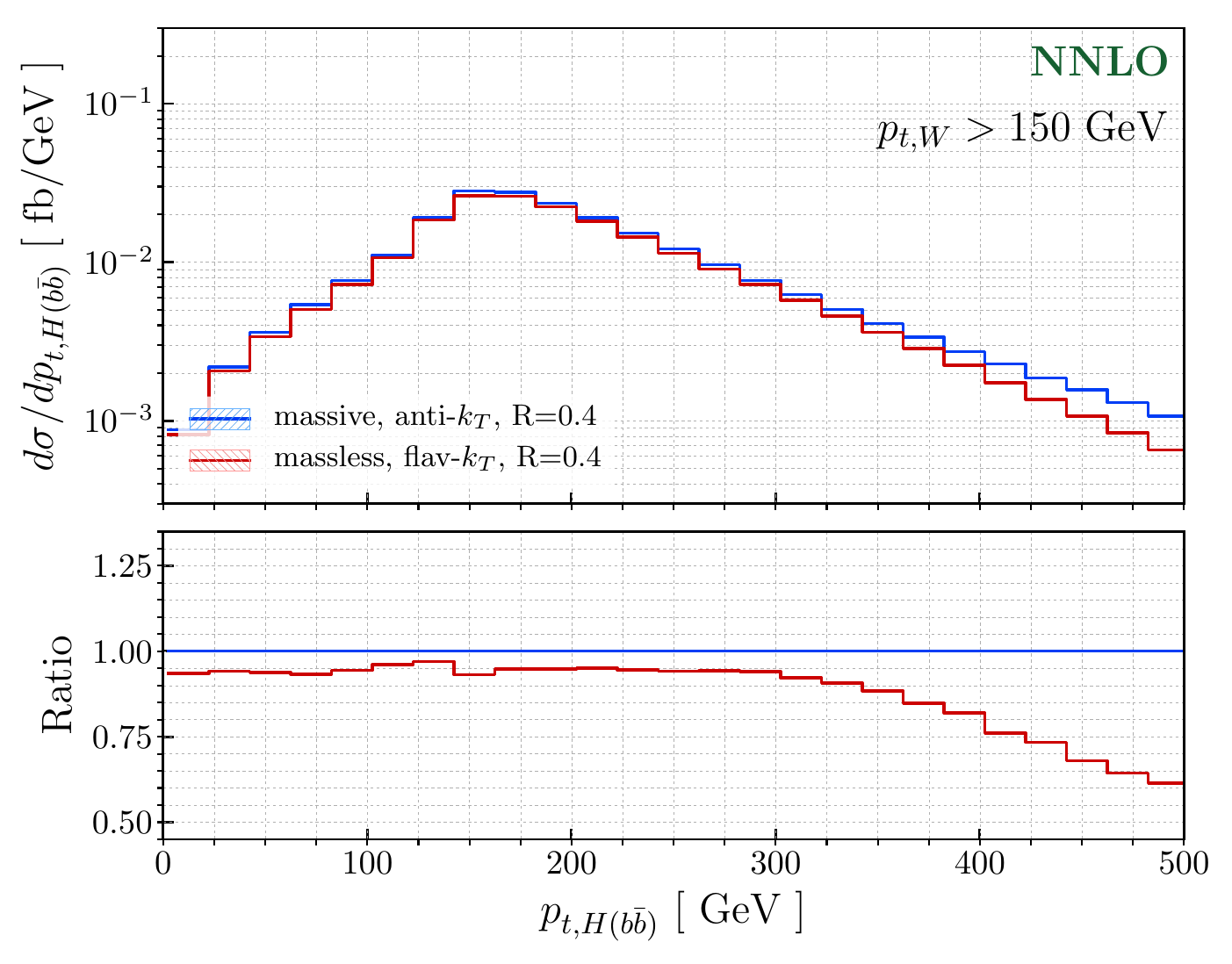}
  \caption{Reconstructed Higgs boson transverse momentum, see text for
    details, calculated at NLO (upper plots) and NNLO (lower plots) for central values of the renormalization and factorization scales.
Lower panes
    show ratios of massless to massive results.
 See text for details.
  }
  \label{fig:ptHbb}
\end{figure}

Next, we consider the transverse momentum distribution of those $b$-jet pairs whose invariant mass is closest to the
mass of the Higgs boson.
The corresponding NLO and NNLO distributions are shown in
Fig.~\ref{fig:ptHbb}.
For standard fiducial cuts and for $\pth \lesssim 300~\gev$, we
observe that  distributions computed  with massive and massless $b$ quarks only differ by
a re-scaling factor whose magnitude follows from the ratios of the fiducial cross sections.
However, for higher transverse momenta, the difference  between massive  and massless  calculations
grows rapidly and becomes  as large as $\mathcal{O}(25\%)$ at about
$\pth \sim 400~\gev$.
This effect is driven by differences in clustering sequences of
the employed jet algorithms and it is present already at leading order.
Indeed, at very high transverse momenta,  decay
products of the Higgs boson are collimated and can be clustered within
a single  jet with zero bottom quantum number. Such events are then rejected by fiducial cuts since
(at least) two $b$ jets are required.    Since such a clustering starts to occur earlier in case of the
flavor-$\kt$ jet algorithm, the massless result falls off
more rapidly  than the massive one. To some extent, this difference can be mitigated
if  a smaller clustering radius for the
flavor-$\kt$ jet algorithm is chosen while the jet radius for the usual anti-$\kt$ algorithm is kept fixed.  We have verified
that such choices  lead to increased values of
$\pth$  at which massive and massless results start to
depart from each other.

\begin{figure}\centering
  \includegraphics[width=0.45\textwidth]{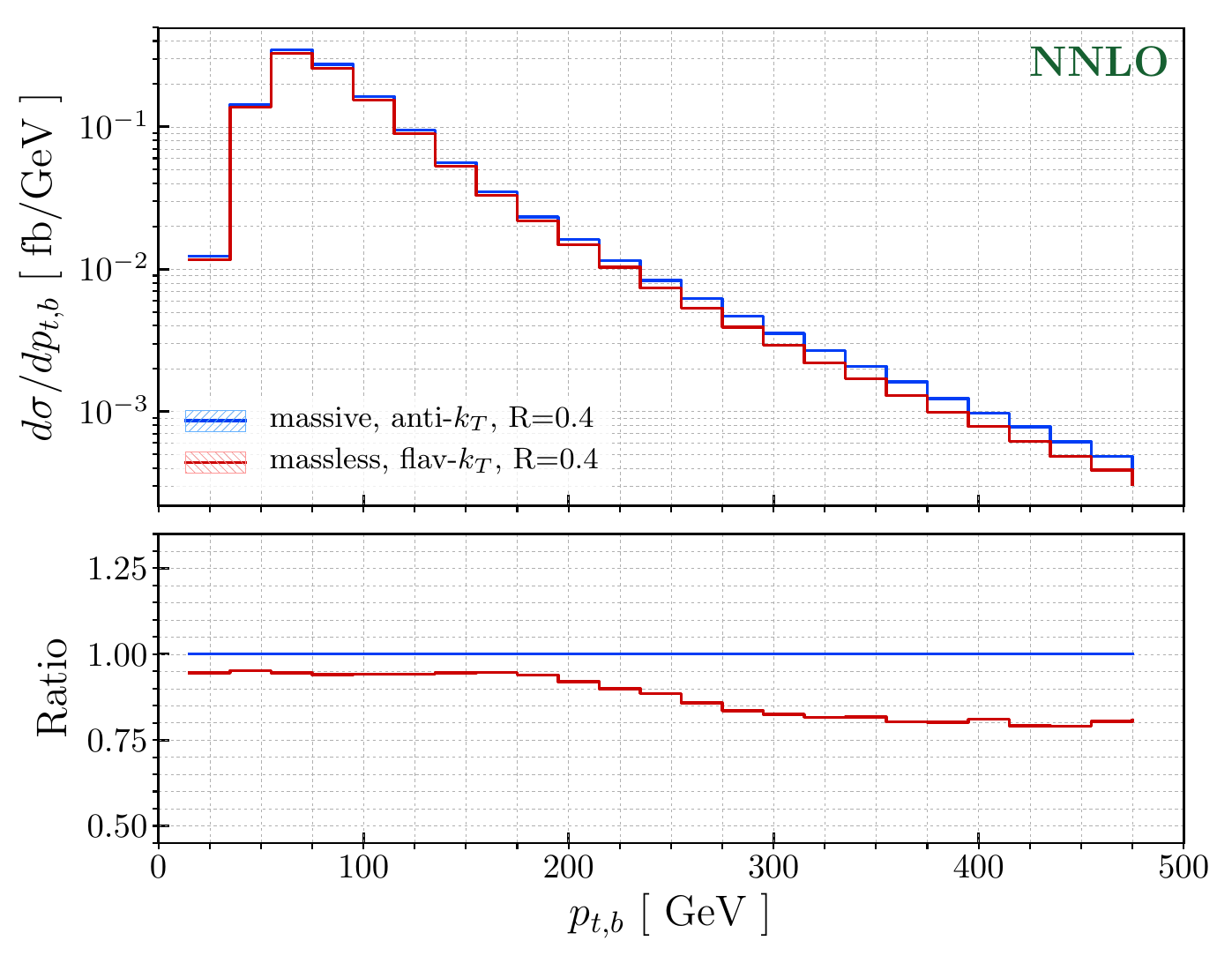}
  \includegraphics[width=0.45\textwidth]{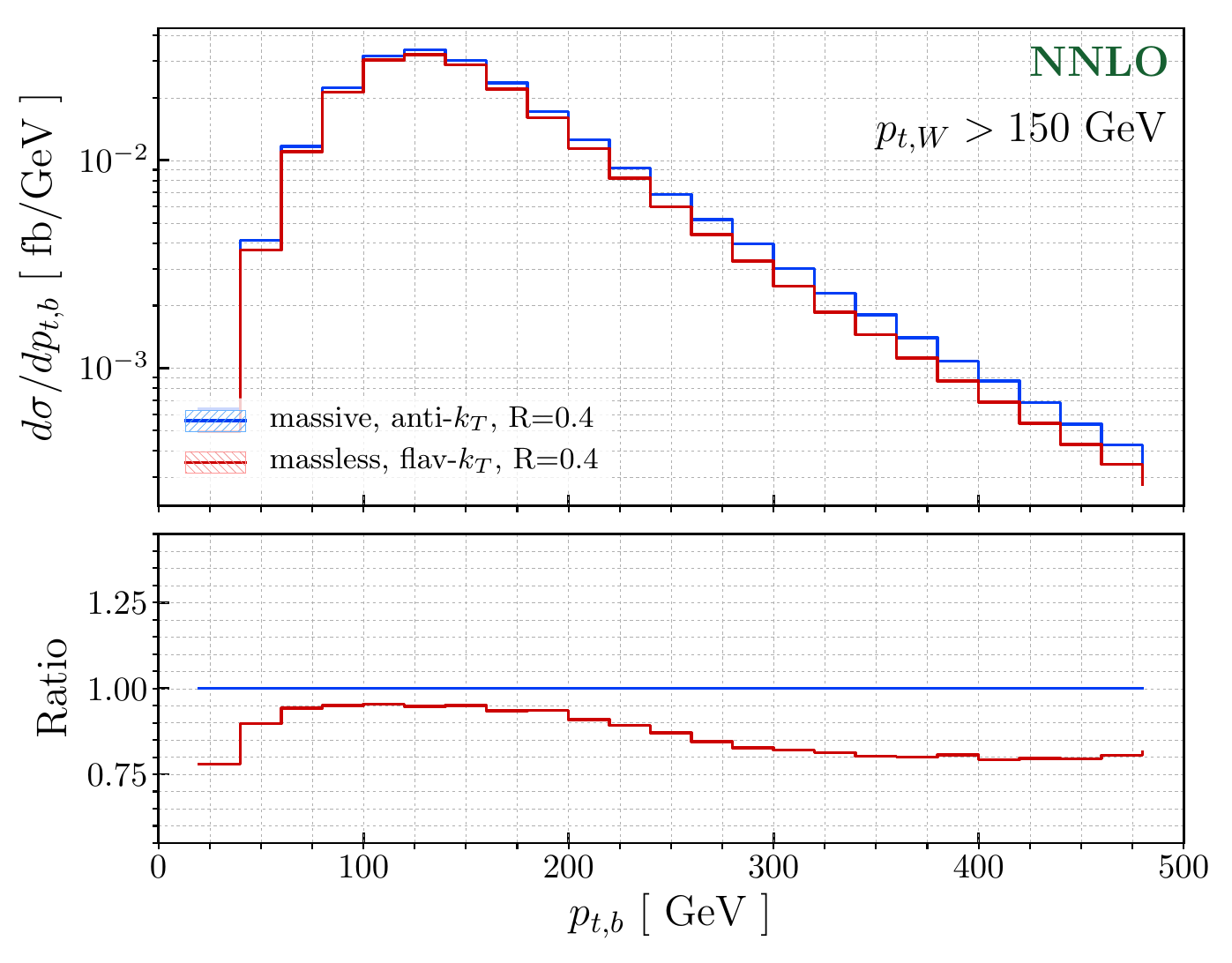}
  \caption{The transverse momentum distribution of the leading $b$ jet
    calculated at NNLO for central values of the renormalization and factorization scales.
Lower panes
    show ratios of massless to massive results.
 See text for details.
  }
  \label{fig:ptbjet}
\end{figure}
\begin{figure}\centering
  \includegraphics[width=0.45\textwidth]{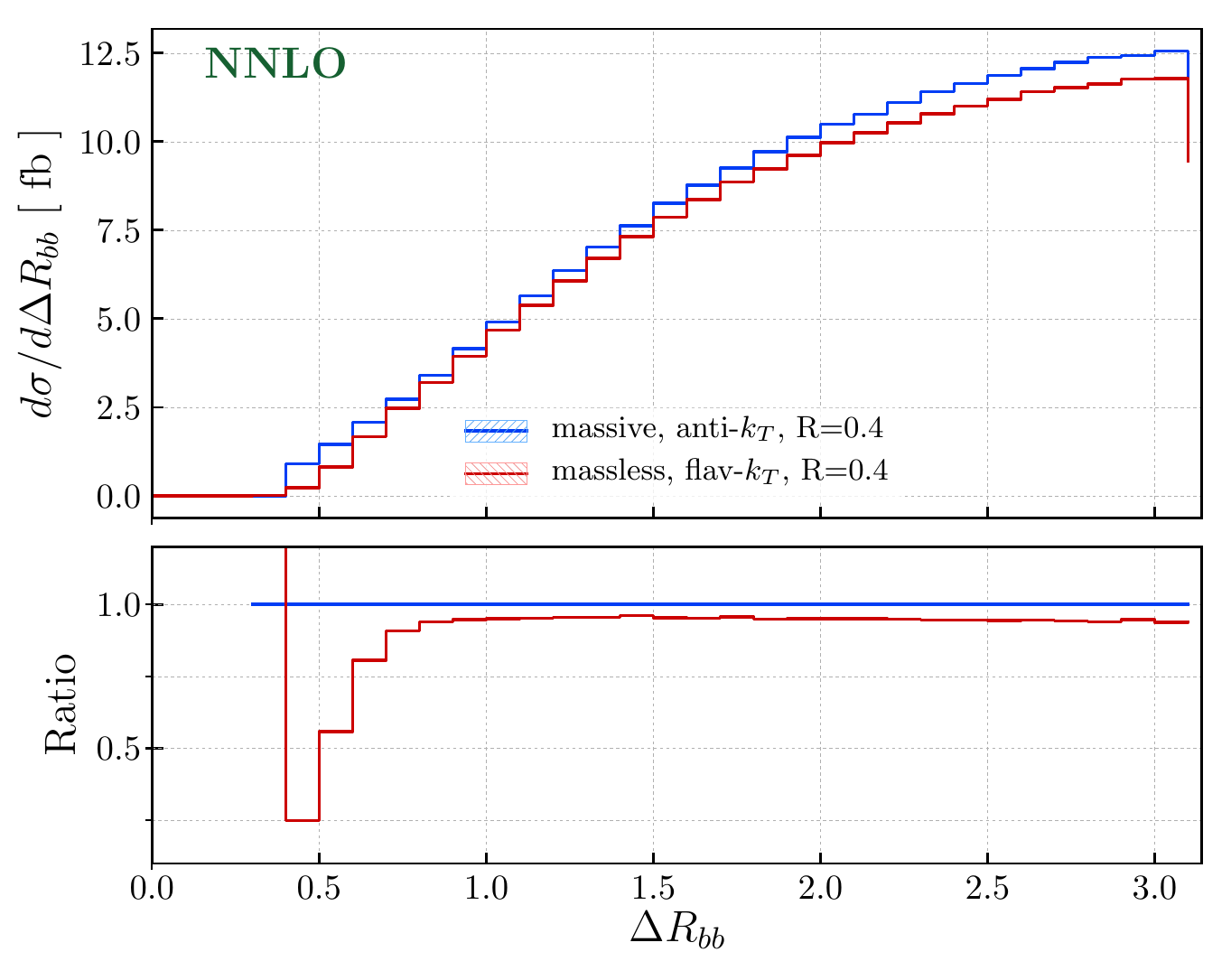}
  \includegraphics[width=0.45\textwidth]{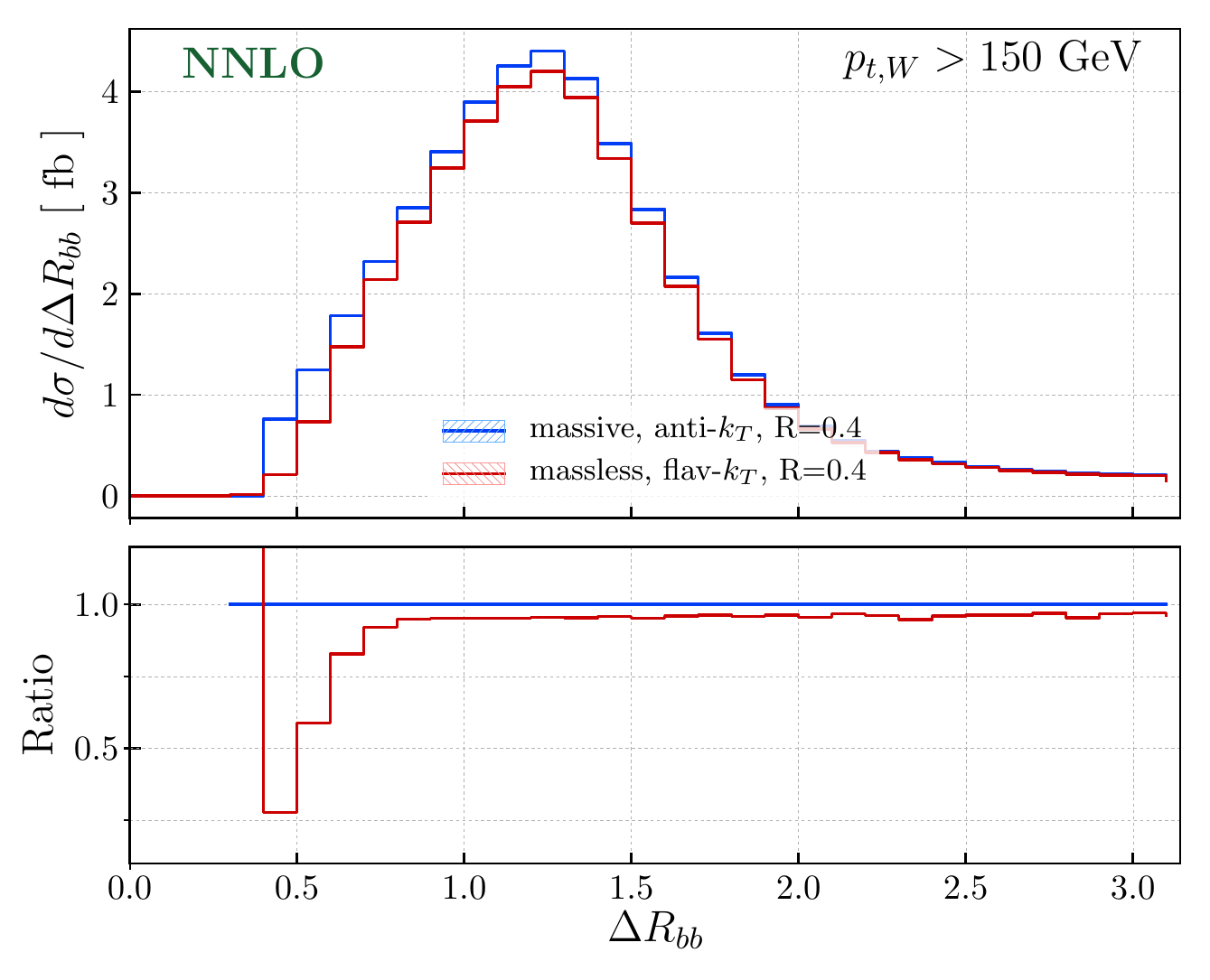}
  \caption{ The distance $\dRbb$ between the two $b$ jets
    used for Higgs boson reconstruction calculated at NNLO for central values of the renormalization and factorization scales.
Lower panes
    show ratios of massless to massive results.
 See text for details.
  }
  \label{fig:Rbb}
\end{figure}
Finally, we show the transverse-momentum distribution of the leading
$b$ jet in Fig.~\ref{fig:ptbjet} and the angular distance
between the two
$b$ jets $\dRbb$  in Fig.~\ref{fig:Rbb}.
We observe significant  differences between massive and massless results at large values of $\ptb$ and
at $\dRbb \sim R$.
Deviations  at large transverse momenta in the $\ptb$ distribution
have the same origin as differences  observed in $\pth$ distributions. As we discussed earlier,
they are related to differences in the clustering of two $b$ jets into a single jet in the massive and massless cases.

In case of the $\dRbb$ distributions, the massless to massive ratio is
flat for large $\dRbb \gtrsim 0.75$ jet separation but they become
different for smaller values of $\dRbb$.  Again, these features are
closely related to the behavior of the $\pth$ distributions since a small
angular separation of the two $b$ jets corresponds to a boosted
configuration from a Higgs boson with large transverse momentum.

\begin{figure}\centering
  \includegraphics[width=0.45\textwidth]{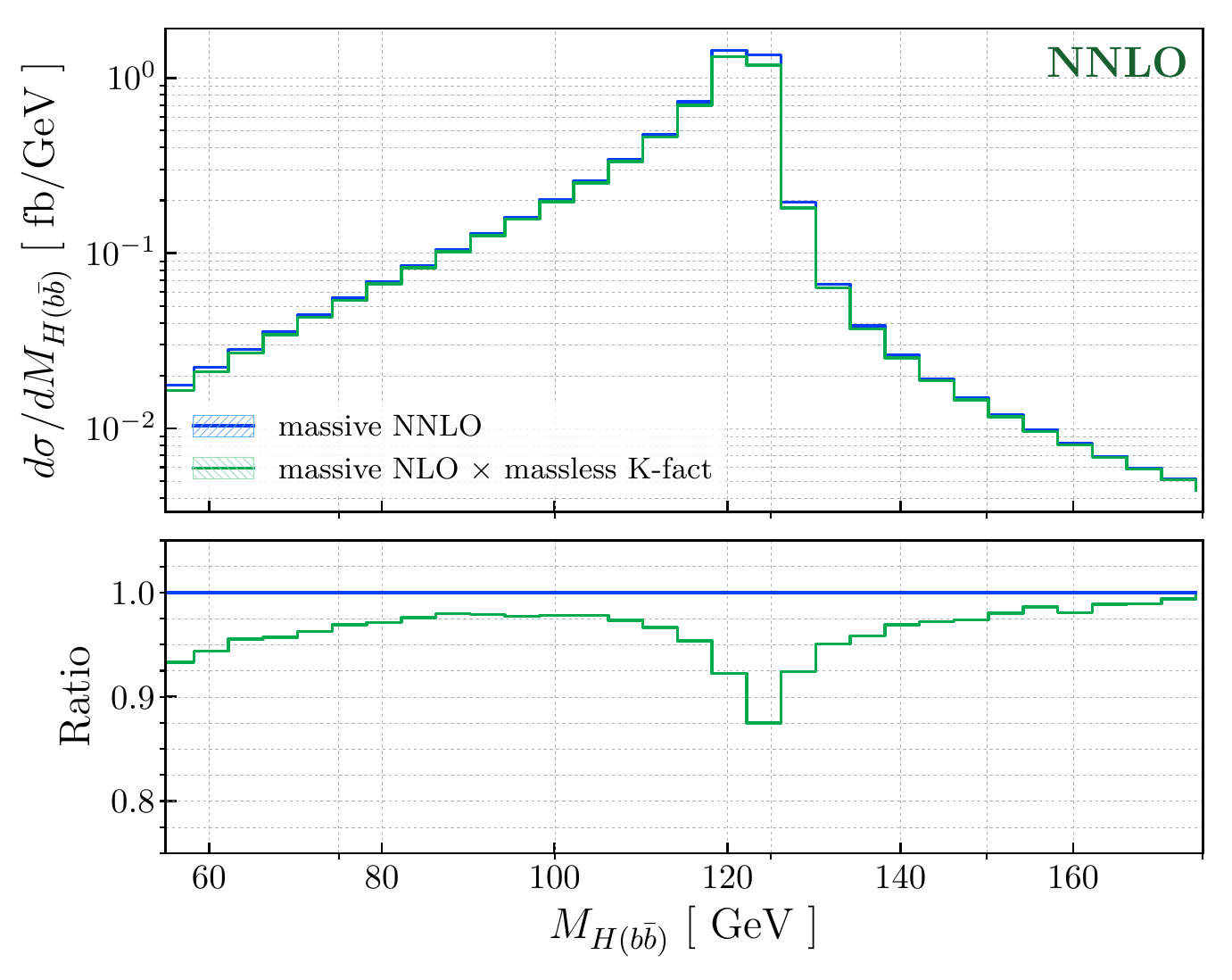}
  \includegraphics[width=0.45\textwidth]{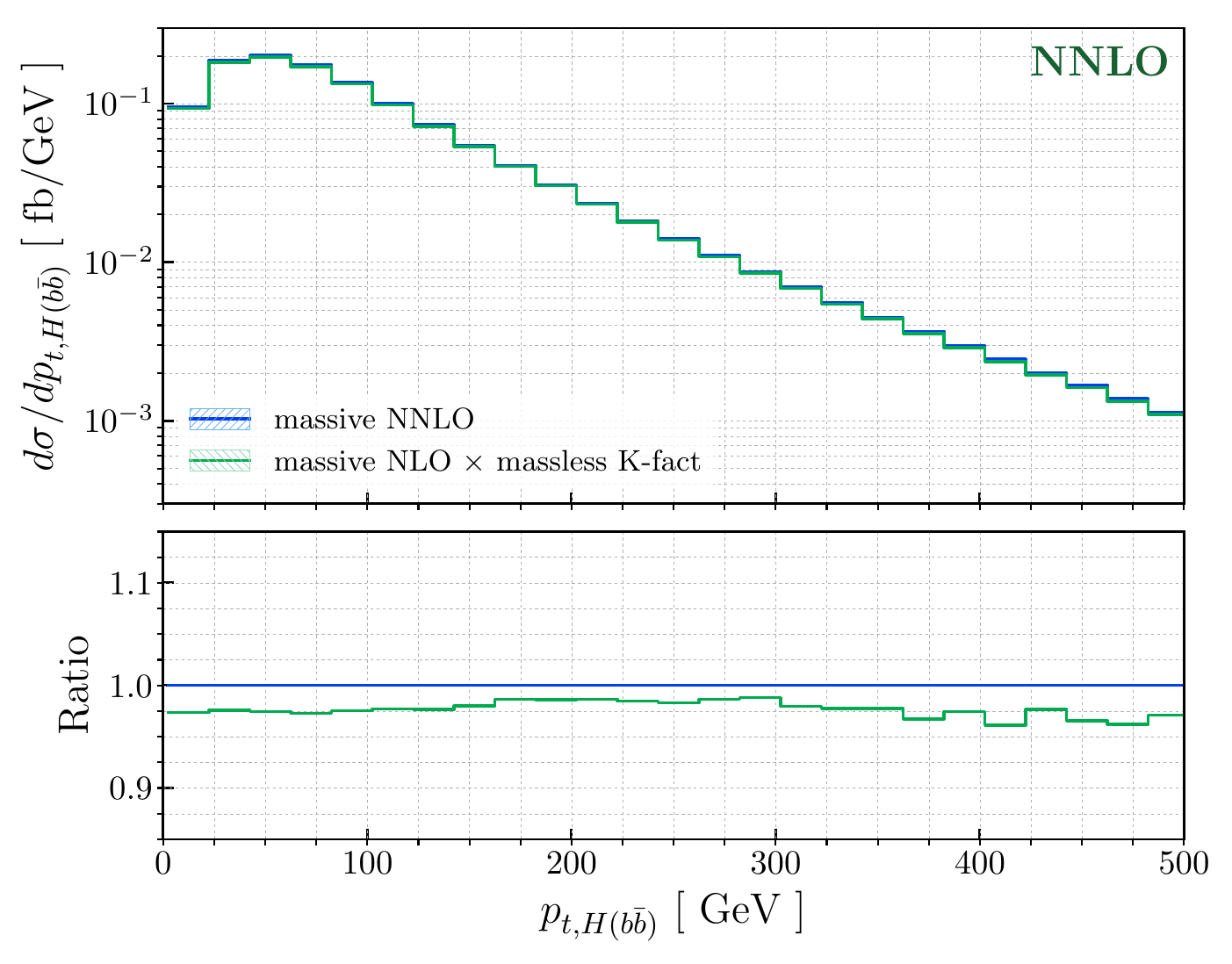}
  \caption{Comparison of approximate and exact NNLO distributions for
    central values of the renormalization and the factorization
    scales. Lower panes show ratios of the two distributions with
    respect to the exact calculation.  See text for details.  }
  \label{fig:kfactors}
\end{figure}

As we already pointed out, some differences in kinematic distributions
computed with massive and massless quarks arise already at leading
order. If we assume that radiative effects are similar in massive and
massless cases, one can construct approximate NNLO distributions from
massive NLO computations and massless differential $K$-factors defined
as ${\rm d} K = {\rm d} \sigma_{\NNLO}/{\rm d} \sigma_{\NLO}$.
We compare the (so constructed) approximate and exact NNLO
distributions for $\mbb$ and $\pth$ in
Fig.~\ref{fig:kfactors}.
We observe that such an approximation is only partially successful; it
provides a decent description of the true $\pth$ distribution but
does not capture all the details of the $\mbb$ spectrum.

\section{Comparison with parton shower}
\label{sect:shower}

Having discussed  fixed-order calculations with
massive and  massless $b$ quarks,  we turn to a comparison of these calculations
with  parton showers.
Such a comparison is important because   experimental analyses often
rely on parton showers and one needs  to understand their reliability
by comparing them to fixed-order computations.

For our purposes, we use the \texttt{POWHEG-BOX-V2}
framework~\cite{Nason:2004rx,Frixione:2007vw,Alioli:2010xd} with the
publicly available \texttt{HWJ} event generator~\cite{Luisoni:2013kna}
constructed using the improved \texttt{MiNLO}
method~\cite{Hamilton:2012rf}. It allows us to simulate the
$pp \to W^{+}Hj$ process with NLO QCD accuracy. Moreover, upon
integration over the resolved radiation, the NLO QCD result for
$pp \to W^{+}H$ is obtained.
For the parton shower we use
\texttt{Pythia8}~\cite{Sjostrand:2007gs} with the Monash
tune~\cite{Skands:2014pea}.
We simulate the $H\to\bb$ decay with \texttt{Pythia8} that includes
the matrix element correction that allows to describe $H \to \bb g$
decay in a reliable way.
To stay  as close as possible to  fixed-order calculations, we
use parton-shower results at the parton level, without hadronization
and multi-parton interactions effects.

Using the \texttt{POWHEG+Pythia8} setup\footnote{Note that we use the
  ``out-of-the-box'' implementation of \texttt{HWJ} process which, at
  variance to our NNLO calculation, includes off-shell $W$ bosons and
  the physical CKM matrix.}  and our fiducial cuts described in
Sec.~\ref{sect:WHprod}, we obtain the following values for the cross
sections
\begin{align}
  \label{eq1}
  \sigma_{\rm fid}^{\texttt{PWHG+Pythia8}}
  ={}&
       23.934(9)~\fb\,,
       &
  \sigma_{\rm fid, boost}^{\texttt{PWHG+Pythia8}}
  ={}&
       4.368(4)~\fb\,.
\end{align}
The second  result shown in  Eq.~\eqref{eq1}  is obtained by requiring that, in addition to standard fiducial cuts,
the  transverse momentum of the $W$ boson $\ptw$  exceeds $150~{\rm GeV}$. The
uncertainties shown in Eq.~\eqref{eq1}  correspond to numerical integration errors.

The parton-shower cross sections  Eq.~\eqref{eq1} differ from  NNLO cross sections computed  with massive
$b$ quarks by about $2\%$ in
the full fiducial region and by about $4\%$ if the additional $\ptw$ cut
is applied (c.f.\ Table~\ref{tab:fid-xsec}).
These differences are only natural given that  the \texttt{POWHEG+Pythia8+MiNLO} setup
is different compared to what we use to obtain fixed-order predictions, see Ref.~\cite{Luisoni:2013kna} for further
details.

\begin{figure}\centering
  \includegraphics[width=0.45\textwidth]{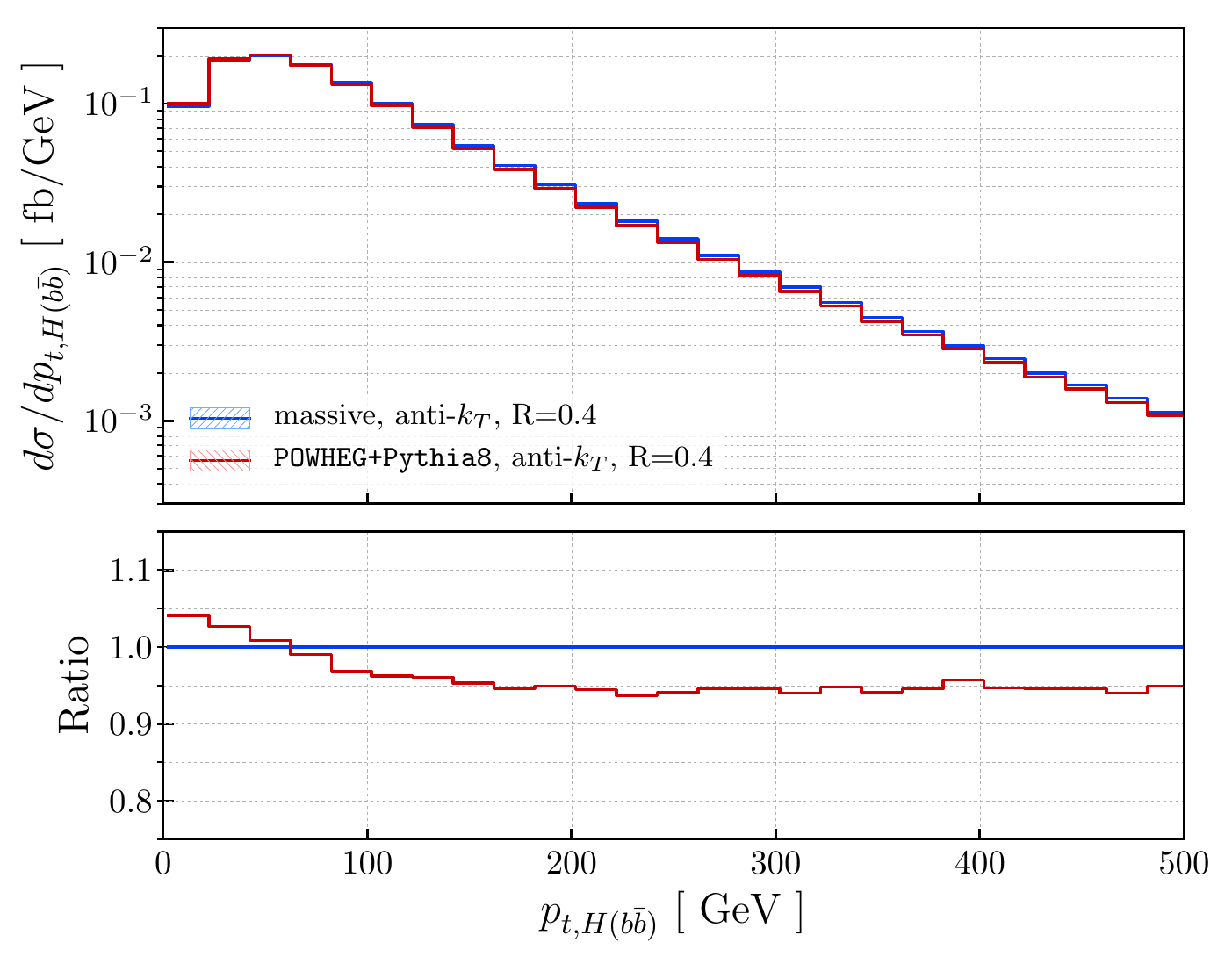}
  \includegraphics[width=0.45\textwidth]{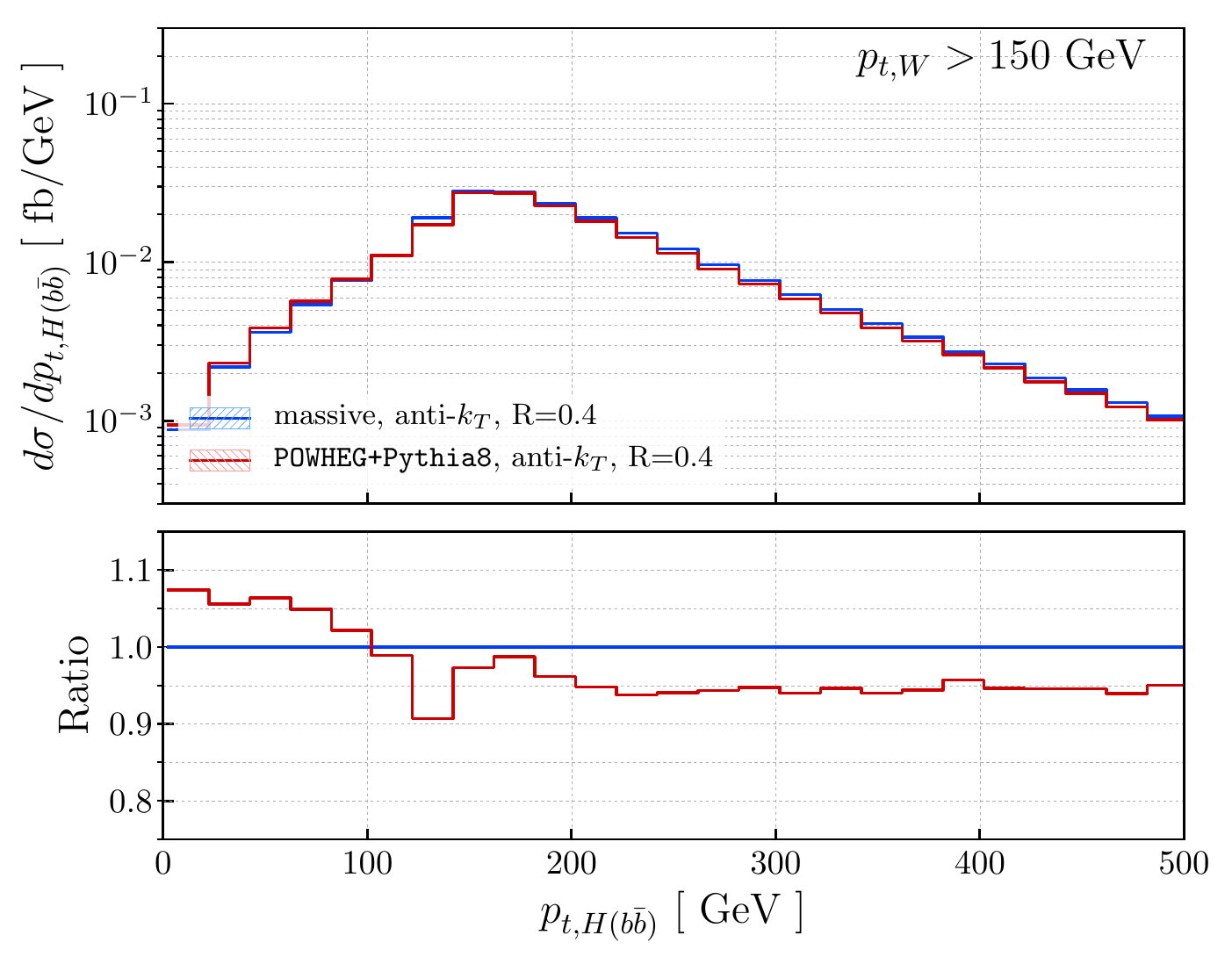}
  \caption{ The transverse momentum distribution of two $b$ jets whose
    invariant mass is closest to the Higgs boson mass for central
    values of the renormalization and factorization scales.  Lower
    panes show ratios of parton shower to massive fixed-order results.
    See text for details.  }
  \label{fig:pwhg-pth}
\end{figure}
\begin{figure}\centering
  \includegraphics[width=0.45\textwidth]{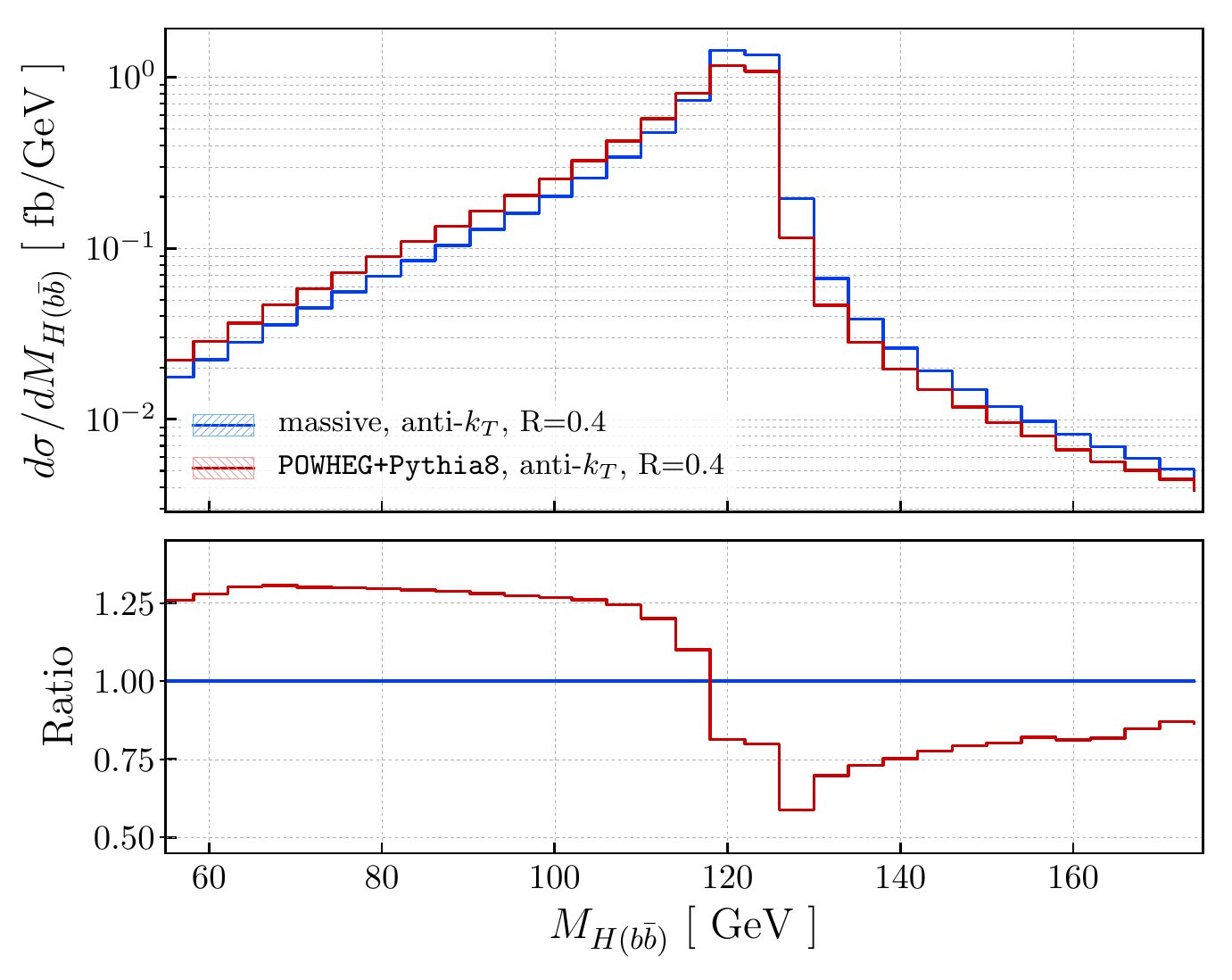}
  \includegraphics[width=0.45\textwidth]{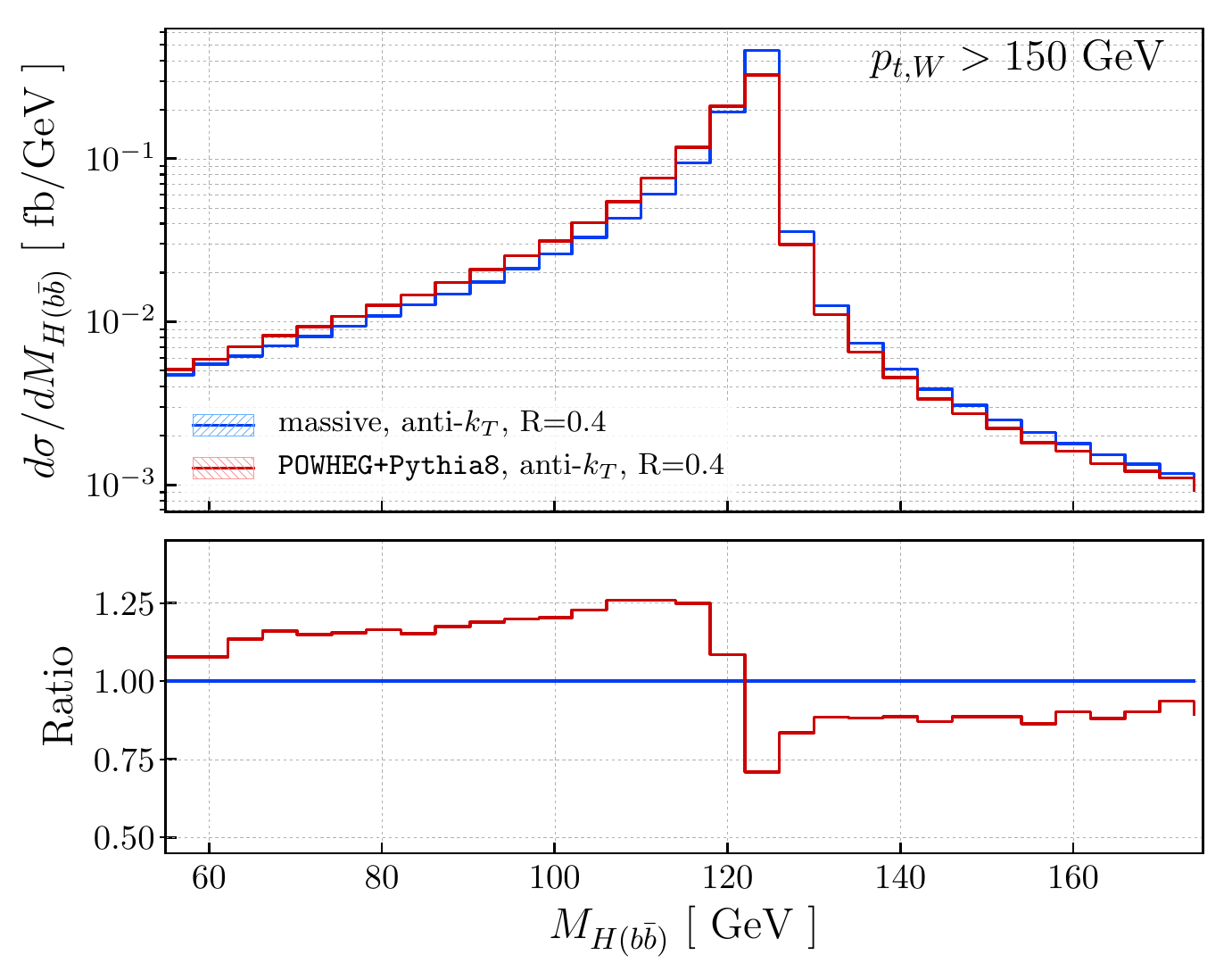}
  \caption{ The invariant mass distribution of the two $b$ jets that
    best reconstruct the Higgs boson mass for central values of the renormalization and factorization scales.
  Lower panes
    show ratios of parton shower to massive fixed-order results.
 See text for details. }
  \label{fig:pwhg-mbb}
\end{figure}

We proceed with  the comparison of  fixed-order
and the parton-shower descriptions of selected kinematic distributions
for a pair of $b$ jets whose invariant mass is closest to the mass of the Higgs boson.
We present the transverse momentum distribution of such
$b$-jet pairs  in Fig.~\ref{fig:pwhg-pth}, and their invariant mass distribution
in  Fig.~\ref{fig:pwhg-mbb}.
In the case of the transverse momentum distribution, both with and
without the additional $\ptw$ cut, we see that in the region $\pth
\gtrsim 100~\gev$ the parton-shower result is smaller than the massive
NNLO result by about five percent, whereas for transverse momenta
below the peak of the distribution, $\pth \lesssim 50~\gev$, the
parton-shower prediction exceeds the fixed-order result by about five
percent.  We note that such behavior is expected since additional
QCD radiation, simulated by a parton shower, reduces energies of the $b$
jets leading to a softer spectrum.

Differences between parton-shower predictions and the massive
fixed-order NNLO result for the invariant mass of the $\bb$-system are
more significant than in case of the transverse momentum distribution,
c.f. Fig.~\ref{fig:pwhg-mbb}.  Below the Higgs peak we observe a
$\mathcal{O}(25\%)$ excess of the parton-shower result over the
fixed-order result; above the peak, parton-shower results are
$\mathcal{O}(25\%)$ smaller than fixed-order results.
We note that the parton-shower and the fixed-order distributions can
be made well aligned provided that the fixed-order distribution
is shifted along the $x$ axis by $\delta M_{\bb} \sim -4~{\rm GeV}$.

\section{Conclusions}
\label{sect:concl}

In this paper, we discussed the associated production of the Higgs
boson, $pp \to WH$, and the decay of the Higgs boson to $\bb$ pairs at
the LHC.  We included the NNLO QCD corrections to the production and
decay processes, retaining the dependence on the $b$-quark mass.  The
inclusion of the $b$-quark mass in the calculation is important as it
allows us to use realistic jet algorithms to describe $b$ jets, making
theoretical and experimental analyses more aligned.

We compared theoretical predictions
for the associated production that are obtained  with massive and massless $b$ quarks.
 We observed  ${\cal O}(6\%)$ differences between the two results  once
 fiducial cuts are applied. Such relatively large   differences can be traced back to different
 acceptances in radiative decays of the Higgs boson  $H\to\bb g$  when they are  computed in the massive
 and in the massless approximations for a standard set of fiducial volume cuts.
We also found that  radiative corrections to the production
process are less sensitive to $b$-quark mass effects.

Interestingly, mass effects can become much more pronounced in
kinematic distributions.
For example, we observed large differences between massive and
massless predictions in kinematic regions where $b$ jets have large
transverse momenta.  In these cases, differences in clustering
algorithms employed with massive and massless partons, needed to
unambiguously define a jet's flavor, combine with rapidly changing
distributions and lead to ${\cal O}(20\%)$ discrepancies between
the theoretical predictions.

We note that in some cases such large discrepancies are driven by
differences in lower-order distributions while massive and massless
$K$-factors turn out to be similar. If this is the case, an
approximate massive NNLO result may be constructed from massive NLO
result and massless NNLO/NLO $K$-factor. We have identified the
transverse momentum $\pth$ as one such observable. However, there are
also other cases where the differences in NNLO distributions are
driven by different (massive and massless) $K$-factors; if this is the
case, the approximate distribution will not provide a decent
description of the true result. This is the case, e.g., for the invariant
mass $\mbb$.

Differences between massive NNLO QCD and parton-shower computations,
discussed in Sec.~\ref{sect:shower}, are easily understood if we
assume that parent $b$ quarks lose more energy in a parton-shower
computation than in a fixed-order one.  This implies that shapes of,
at least some, distributions in both cases are similar but the
distributions themselves are shifted relative to each other, e.g.
${\rm d} \sigma^{(PS)}/{\rm d}x \; (x ) \sim {\rm d}
\sigma^{(FO)}/{\rm d} x \; (x + \delta_x)$. We have found that, in
case of the invariant mass of two hardest $b$ jets, $\delta_x \sim
4~{\rm GeV}$ which appears to be a rather natural value.

In summary, we studied effects of the $b$-quark mass on associated
production of the Higgs boson, $pp \to WH$, followed by decay of the
Higgs boson into a $\bb$ pair. Although such effects are not large, we
found that they are typically larger than naively expected and that
they can affect both fiducial cross sections and kinematic
distributions in a somewhat unexpected way. We look forward to future
studies of such effects in other processes relevant for the LHC
phenomenology.

{\bf Acknowledgments:}
We would like to thank Gavin Salam for useful discussions as well as
providing us with his private implementation of the flavor-$\kt$
algorithm~\cite{Banfi:2006hf}.
This research is partially supported by BMBF grant 05H18VKCC1 and by
the Deutsche Forschungsgemeinschaft (DFG, German Research Foundation)
under grant 396021762 - TRR 257. The research of F.C. was partially
supported by the ERC Starting Grant 804394 hipQCD.

\appendix
\section{Renormalization}
\label{app:renorm}
In this appendix, we discuss the details of the renormalization scheme
that we adopt in this calculation. We work with $\nf=4$ active flavors
in the proton, but we renormalize the strong coupling constant
$\alpha_s$ with $\nf=5$, in the $\MSbar$ scheme. As was already
mentioned in the main text we renormalize the $b$-quark mass $m_b$ on
the mass shell, but use the $\MSbar$ mass at the scale $M_H$ in the
calculation of the bottom Yukawa coupling that enters the Higgs decay
rate computation.

The renormalization of the $H\to \bb$ decay process was discussed at
length in Ref.~\cite{Behring:2019oci} and we do not repeat it here.
Instead, in this appendix, we focus on the production process.  We
start by discussing the renormalization of the $q\bar q \to VH $
amplitude $\mathcal{A}$ with $q$ being a massless quark, i.e.
$q \ne b$.
Neglecting $b$-quark contributions altogether and considering $\nf=4$
massless flavors, we write the $\overline{\rm MS}$-renormalized
amplitude as
\begin{align}
  \mathcal A^{(\nf=4)} = \mathcal A_0 +
  \left(\frac{\alpha_s^{(4)}}{2\pi}\right) \mathcal A_1 +
  \left(\frac{\alpha_s^{(4)}}{2\pi}\right)^2 \mathcal A_2^{(\nf=4)} +
  \mathcal O(\as^3)
  \,,
  \label{eq:a1}
\end{align}
where by $\alpha_s^{(\nf)}$ we denote the $\MSbar$-renormalized strong
coupling constant defined in a theory with $\nf$ massless flavors and
evaluated at a scale $\mu$.
We note that an explicit dependence on the number of active flavors
appears in the renormalized amplitude only at the 2-loop level,
cf. Eq.~(\ref{eq:a1}).

\begin{figure}
  \centering
  \begin{subfigure}[b]{0.31\textwidth}
    \centering
    \includegraphics[width=0.5\textwidth]{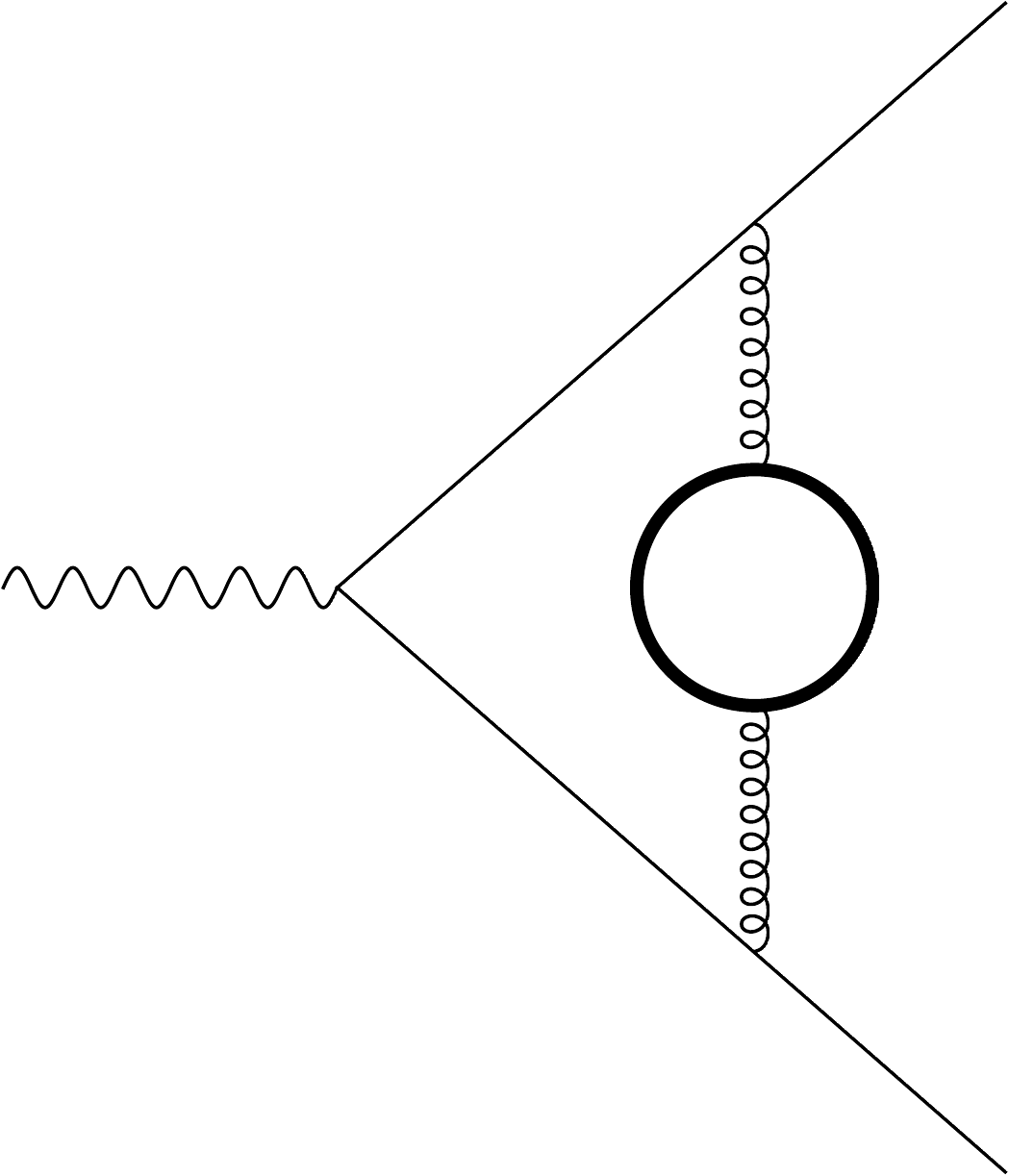}
    \caption{Electromagnetic vertex}
    \label{fig:bvert}
  \end{subfigure}
  \begin{subfigure}[b]{0.31\textwidth}
    \centering
    \includegraphics[width=0.8\textwidth]{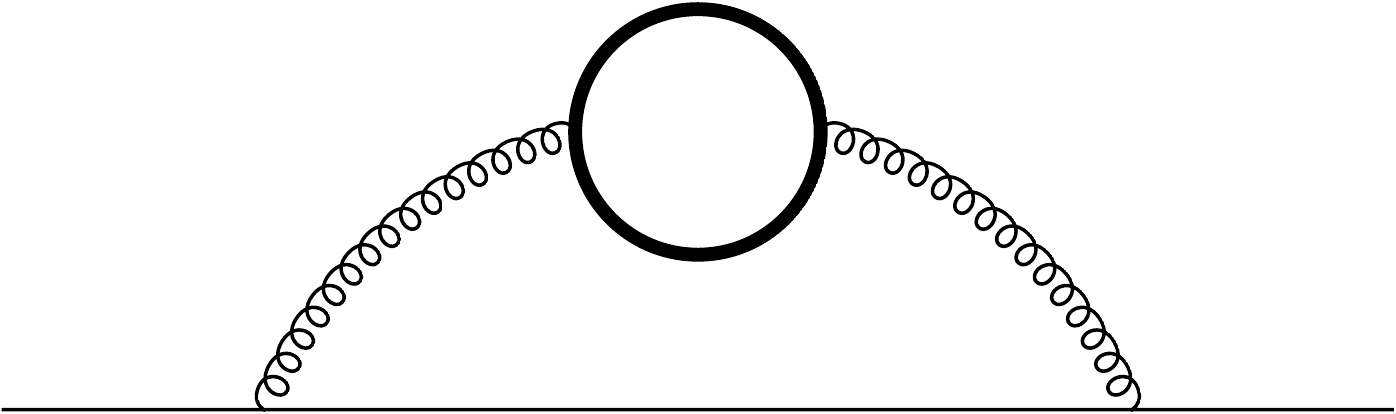}
    \vspace{1.0cm}
    \caption{Light-quark self-energy}
    \label{fig:b2loop}
  \end{subfigure}
  \begin{subfigure}[b]{0.31\textwidth}
    \centering
    \includegraphics[width=0.8\textwidth]{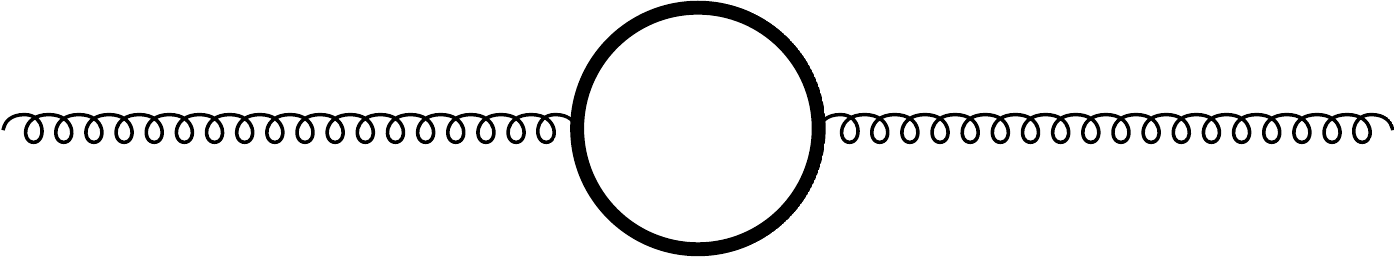}
    \vspace{1.0cm}
    \caption{Gluon self-energy}
    \label{fig:g1loop}
  \end{subfigure}
  \caption{The $b$-quark contribution to the electromagnetic vertex
    (a) and to the light quark self-energy (b). In both cases,
    corrections can be expressed in terms of the $b$-contribution to
    the gluon self-energy (c). In this figure, massive quarks are
    denoted by a thick line.  See text for details.  }
  \label{fig:vertex}
\end{figure}
We continue by expressing  Eq.~\eqref{eq:a1} through the bare
coupling constant $\asb$ and find
\begin{align}
  \mathcal A^{(\nf=4)}
  ={}&
       \mathcal{A}_0
       + \asbontwopi \mathcal{A}_1
       + \asbontwopi^2 \left\{
       \mathcal{A}_2^{(\nf=4)} + \frac{\beta_0^{(\nf=4)}}{\ep} \mathcal{A}_1
       \right\} +
       \mathcal O(\asb^3),
\label{eq:a2}
\end{align}
where $S_\ep = (4\pi)^\ep e^{-\ep \gamma_E}$ is the standard
$\MSbar$ factor and
\begin{align}
  \beta_0^{(\nf)} ={}& \frac{11}{6}\Ca -\frac{2}{3}\tr \nf\,.
\end{align}
In order to include the $b$-quark contribution to Eq.~\eqref{eq:a1},
we need to add the gluon vacuum polarization diagram
Fig.~\ref{fig:bvert} and to account for additional contributions to
renormalization constants that arise in the theory due to loops
with massive $b$ quarks.
For the amplitude ${\cal A}$ an additional renormalization factor is
the wave function renormalization constant of a massless quark $Z_q$
that receives $b$-quark contributions at two loops, see
Fig.~\ref{fig:b2loop}.
Another contribution that arises in the theory with massive $b$ quarks
is the gluon wave function renormalization constant $Z_A$,
see Fig.~\ref{fig:g1loop}.

Starting from Eq.~\eqref{eq:a2}, we re-express the renormalized
amplitude through the coupling constant defined in a theory with five
active flavors.  We find
\be
\begin{split}
  \mathcal A^{(\nf=5)}
  ={} Z_{q} \Bigg \{  \mathcal A_0 & +
\left(\frac{\as^{(5)}}{2\pi}\right)\mathcal A_1
+
\left(\frac{\as^{(5)}}{2\pi}\right)^2
\left[\mathcal A_2^{(\nf=4)} + \frac{1}{\ep} \lp \beta_0^{(\nf=4)}-\beta_0^{(\nf=5)} \rp
\mathcal A_1\right]
\\
&
+
\left(\frac{\as^{(5)}}{2\pi}\right)^2 \mathcal A_2^{(b,\rm bare)}+
\mathcal O(\as^3) \Bigg \}.
\label{eq:app_nf5}
\end{split}
\ee
From now on, we will always work with $\as$ renormalized in a theory
with $\nf=5$ massless flavors at a scale $\mu$. Therefore, unless
stated otherwise, we will use the short-hand notation
$\as = \as^{(5)}(\mu)$.

To proceed further, it is convenient to express $\mathcal A^{(\nf=5)}$
through two-loop contributions to the wave function renormalization
constants $Z_{q}$ and $Z_A$.
To this end, we write
\begin{align}
  \begin{aligned}
    Z_q ={}& 1 + \asbontwopi^2 \tilde\Sigma_2(0)+ \mathcal O(\as^3)\,, \\
    Z_A ={}& 1 - \asbontwopi \Pi_1(0)+ \mathcal O(\as^2)\,.
  \end{aligned}
\label{eq:self}
\end{align}
We leave the discussion of the massless quark and gluon self-energies,
$\tilde\Sigma_2(0)$ and $\Pi_1(0)$, to Appendix~\ref{app:formf}.
Here, we only remark that the difference of the two
$\beta$-functions in Eq.~\eqref{eq:app_nf5} can be expressed through
$\Pi_1(0)$ and an additional constant term, cf.
Eq.~\eqref{eq:pi0}.
Hence, we write
\begin{align}
  \frac{1}{\ep} \lp \beta_0^{(\nf=4)}-\beta_0^{(\nf=5)}
  \rp = \Pi_1(0) + K_1,
\label{eq:dec}
\end{align}
with $K_1 \equiv \frac{2}{3}\tr\ln\lp m_b^2/\mu^2 \rp + {\cal O}(\ep)$.

Using Eqs.~\eqref{eq:dec} and~\eqref{eq:self} we write
Eq.~\eqref{eq:app_nf5} as
\begin{align}
  \begin{aligned}
    \mathcal A^{(\nf=5)} = \mathcal A_0 + \asontwopi \mathcal A_1 +
    \asontwopi^2\left\{\mathcal A_2^{(\nf=4)} + K_1 \mathcal A_1+
      \mathcal A_2^{b\rm,reg}\right\}+ \mathcal O(\as^3),
  \end{aligned}
\end{align}
where we introduced
\begin{align}
\mathcal A_2^{b\rm,reg} =
\mathcal A_2^{b\rm,bare}+\tilde \Sigma_2(0)\mathcal A_0 +
\Pi_1(0) \mathcal A_1.
\label{eq:ab2reg}
\end{align}
The square of the amplitude $\mathcal{A}^{(\nf=5)}$ expanded to second
order in $\alpha_s$ gives the following contribution to the cross
section
\begin{align}
  \begin{aligned}
    \int \left|\mathcal A^{(\nf=5)}\right|^2 \; {\rm dLips}
    \sim{}&
    \dsig{\LO}
    + \asontwopi \dsig{\rm V}
    + \asontwopi^2 \lp
    \dsig{{\rm VV},(\nf=4)}
    + K_1 \dsig{\rm V}
    + \dsig{{\rm VV},(b,\rm reg)}
    \rp,
\end{aligned}
\label{eqa10}
\end{align}
where $\dsig{\rm V}$ and $\dsig{\rm VV}$ are the one- and the two-loop
contributions to cross sections, respectively, and
$\dsig{{\rm VV},(b,\rm reg)}$ is the two-loop contribution
proportional to
$2{\rm Re}\left( \mathcal A_0^{\dagger} A_2^{(b,\rm reg)}\right)$.
For completeness, we report the explicit result for
$\mathcal A_2^{(b,\rm reg)}$ in Appendix~\ref{app:formf}.

We now discuss the amplitude $q \bar q \to VH+g$ which is needed to
describe real and real-virtual contributions to NLO and NNLO cross
sections.
As in the previous case, we start with the amplitude computed in a
theory with $\nf=4$ massless quarks and write
\begin{align}
  \mathcal A_j^{(\nf=4)}
  ={}&
       g_s^{(4)}
       \left\{\mathcal A_{0,j}
       + \left(\frac{\as^{(4)}}{2\pi}\right) \mathcal A_{1,j}^{(\nf=4)}
       + \mathcal O(\as^2)\right\}.
       \label{eq:qqWHg:4flav}
\end{align}
In  Eq.~\eqref{eq:qqWHg:4flav}    $g_s^{(4)}$ stands for the strong coupling constant in the theory with
four massless flavors, $g_s^{(4)} = \sqrt{4\pi\alpha_s^{(4)}}$.
Equivalently, we re-express Eq.~\eqref{eq:qqWHg:4flav} using the bare coupling constant
\begin{align}
  \mathcal A_j^{(\nf=4)}
  ={}& g_{s,b} \sqrt{S_\ep}
       \left\{\mathcal A_{0,j}
  + \asbontwopi \left[\mathcal
    A_{1,j}^{(\nf=4)}+\frac{\beta_0^{(\nf=4)}}{2\ep} \mathcal A_{0,j} \right] + \mathcal
  O(\as^2)\right\}.
\end{align}
In this case, there are no explicit $\nf$-dependent contributions to
the unrenormalized amplitude so that all the $b$-quark effects only enter
through the renormalization.
Since $Z_q = 1 + \mathcal O(\as^2)$, we only need to renormalize the
strong coupling constant $\as$ and to multiply the unrenormalized
amplitude by the gluon renormalization factor $\sqrt{Z_A}$.
We obtain
\begin{align}
  \mathcal A_j^{(\nf=5)} = g_s \left\{
  \mathcal A_{0,j} + \asontwopi \left[\mathcal A_{1,j}^{(\nf=4)} +
    \frac{K_1}{2} \mathcal A_0\right] +\mathcal O(\as^2)\right\},
\label{eq:ajnf5}
\end{align}
where $g_s = g_s^{(5)}(\mu)$ is the strong coupling constant defined in
the theory with five flavors and renormalized at a scale $\mu$.
We finally write the contribution of the renormalized
$q \bar q \to VH + g$ amplitude Eq.~\eqref{eq:ajnf5} to the cross
section
\begin{align}
  \int \left|\mathcal A_j^{(\nf=5)}\right|^2 \; {\rm dLips}
  \sim{}&
          \asontwopi \dsig{\rm R}
          + \asontwopi^2
          \left[ \dsig{{\rm RV},(\nf=4)} + K_1 \dsig{\rm R}
          \right]+...
\label{eqa14}
\end{align}

The last two contributions that we need to discuss are the double-real
emission processes and the PDFs renormalization term.
The double-real emission processes do not require any renormalization
and can be obtained as a direct sum of $n_f=4$ contributions that we
discussed earlier~\cite{Caola:2017dug,Caola:2019nzf} and an additional
{\it finite} contribution where a virtual gluon splits into a massive
$\bb$ pair.

In the context of PDF renormalization, we stress that we work in a
theory with four active massless flavors in the proton, but we write
the result using the QCD coupling constant computed in a theory with
$\nf=5$ flavors.
Taking into account the change in the coupling constant,
\begin{align}
  \alpha_s^{(4)}
  ={}&
       \alpha_s^{(5)}
       \lp
       1 + \asontwopi K_1 + \mathcal O (\as^3)
       \rp,
\end{align}
we find an additional contribution to the NNLO cross
section that reads
\begin{align}
  \dsig{{\rm PDF},(\nf=5)}
  ={}&
       \dsig{{\rm PDF},(\nf=4)}
       + \asontwopi
       \frac{K_1}{\ep}
       \left[
       \hat P^{(0)} \otimes \dsig{\rm LO}
       + \dsig{\rm LO} \otimes \hat P^{(0)}
       \right],
\label{eqa16}
\end{align}
where $\hat P^{(0)}$ are the LO Altarelli-Parisi splitting functions
and ``$\otimes$'' denotes the standard convolution product, see
Ref.~\cite{Caola:2017dug} for more details.

The resulting  NNLO cross section is obtained by combining
Eqs.~\eqref{eqa10},~\eqref{eqa14} and~\eqref{eqa16}. We find that the
terms proportional to $K_1$ assemble themselves into a finite NLO cross
section. Therefore, we write
\begin{align}
  \dsig{{\rm NNLO},(\nf=5)}
  ={}&
       \dsig{{\rm NNLO},(\nf=4)}
       + K_1 \dsig{\rm NLO}
       + \dsig{{\rm VV},(b,\rm reg)}
       + \dsig{{\rm RR},\bb},
\end{align}
where $\dsig{{\rm NNLO},(\nf=4)}$ is the standard $\MSbar$ result in a
theory with $\nf=4$ massless flavors, $K_1$ is the decoupling constant
reported in Eq.~\eqref{eq:dec}, $\dsig{{\rm VV},(b,\rm reg)}$ is the
purely virtual contribution proportional to
$2{\rm Re}\left( \mathcal A_0^{\dagger} A_2^{(b,\rm reg)}\right)$, see
Eq.~\eqref{eq:ab2reg} and $\dsig{{\rm RR},\bb}$ is the contribution of
the real-emission process $q \bar q \to WH + \bb$.
We discuss the calculation of $\dsig{{\rm VV},(b,\rm reg)}$ in
Appendix~\ref{app:formf}.
Finally, we emphasize that no modifications are required
to compute leading and next-to-leading order $WH$ production cross sections.

\section{Contributions of a massive $b$ quark to a two-loop form factor of a massless quark}
\label{app:formf}

In this appendix, we calculate the contribution of a massive $b$ quark
to the two-loop amplitude $\mathcal A_2^{(b,\rm reg)}$ defined in
Eq.~\eqref{eq:ab2reg}. We note that such a calculation was performed in
Refs.~\cite{Kniehl:1989kz,Rijken:1995gi,Blumlein:2016xcy}; we discuss it here for
completeness.

We begin by considering $\mathcal A_2^{(b,\rm bare)}$, which corresponds
to Fig.~\ref{fig:bvert}. Since helicity of a massless quark is
conserved and since flavor-changing currents are anomaly-free, there
is no difference between the form factors of a vector and of a
vector-axial current.
Therefore, for simplicity  we consider radiative corrections to a
matrix element of a  generic vector current
$J^\mu = \bar q \gamma^\mu q$ between the vacuum state and a $q \bar q$
pair $\langle q(p_1) \bar q(p_2)| J_\mu(0) | 0 \rangle$.

We compute the color factors and write the corresponding amplitude as
\begin{align}
  i \mathcal A_2^{(b,\rm bare)}
  ={} i g_s^2 \Cf  \times \int\frac{d^d k}{(2\pi)^d}
\frac{\bar u_1 \gamma_\alpha  \hat k_1
\gamma^\mu \hat k_{2}  \gamma_\beta  v_2}
{k^2 k_1^2 k_{2}^2}
\left[g^{\alpha\beta} -  \frac{k^\alpha k^\beta}{k^2}\right]
\Pi (k^2),
\label{eq:a2bare}
\end{align}
where $k_{1,2} = k \pm p_{1,2}$ and we use the notation $\hat{k} = k_{\mu}\gamma^{\mu}$. $\Pi(k^2)$ is the
${\cal O}(\alpha_s)$ gluon vacuum polarization contribution.  It is
defined through the following equations
\begin{align}
  i \Pi_{\mu\nu}(p)
  ={}&
       -g_s^2 T_R \int \frac{d^d k}{(2\pi)^d}
       \frac
       {
       {\rm Tr}\big[ \gamma_\nu (\hat k + m_b ) \gamma_\mu (\hat k - \hat p + m_b) \big]
       }{
       \big[k^2-m_b^2\big]\big[(k-p)^2-m_b^2\big]
       }\,,
       \label{eq:ipimunu}
\end{align}
\begin{align}
  i \Pi_{\mu\nu}(p^2) ={}& -i  \lp g^{\mu\nu} p^2 - p^{\mu}p^\nu\rp \Pi(p^2)\,, &
  \Pi(p^2) ={}& \asontwopi \Pi_1(p^2).
\end{align}
The gluon self-energy $\Pi(k^2)$ satisfies the once-subtracted dispersion
relation
\begin{align}
\Pi(k^2) = \Pi(0) + \frac{k^2 }{\pi} \int\limits_{4 m_b^2}^{\infty}
\frac{dq^2}{q^2}
\left[\frac{{\rm Im} \left[\Pi(q^2) \right]}{q^2-k^2-i\ep}\right].
\label{eq:disp}
\end{align}
We now insert this dispersion relation into Eq.~\eqref{eq:a2bare}. The
$\Pi(0)$ term gives rise to a contribution proportional to the one loop
amplitude $\mathcal A_1$ in the Landau gauge. However, since $\mathcal A_1$
is gauge-independent, we can write
\begin{align}
\begin{aligned}
i \mathcal A_2^{(b,\rm bare)}  =
& -i \asbontwopi^2 \Pi_1(0) \mathcal A_1
 -
\int\limits_{4m_b^2}^{\infty}\frac{dq^2}{q^2}
           {\rm Im} [\Pi(q^2)]
         \\
& \times \frac{i g_s^2\Cf }{\pi}
\int\frac{d^d k}{(2\pi)^d}
\frac{\bar u_1 \gamma_\alpha \hat k_1
\gamma^\mu \hat k_{2} \gamma_\beta  v_2}
{[k^2-q^2+i0] k_1^2\;k_{2}^2}
\left[g^{\alpha\beta} -  \frac{k^\alpha k^\beta}{k^2}\right].
\end{aligned}
\label{eq:b5}
\end{align}
The term in the second line of Eq.~\eqref{eq:b5} is proportional to
the one-loop vertex correction due to an exchange of a gluon with the
mass $q^2$ in the Landau gauge. As a consequence, it is both UV and
IR finite.  After simple manipulations, we cast Eq.~\eqref{eq:b5} into
the following form
\begin{align}
  i \mathcal A_2^{(b,\rm bare)} + i \asbontwopi^2
  \Pi_1(0) \mathcal A_1 =
  \asbontwopi  \left[\bar u_1 \gamma^\mu v_2\right]
  \frac{\Cf  }{\pi} \int\limits_{4m_b^2}^{\infty}
  \frac{dq^2}{q^2} \;  {\rm Im} \Pi(q^2) \;
  \widetilde{\rm Tri}(d,q^2,s).
  \label{eq:b6}
\end{align}
We note that in the limit $q^2 \to \infty$, both $\Pi(q^2)$ and
$\widetilde{\rm Tri}(d,q^2,s)$ approach constants, so that the
integration over $q^2$ diverges. To remove this divergence, we need to
incorporate the wave function renormalization constant of a light
quark, $(Z_q - 1) \sim \tilde \Sigma_2(0)$, cf Eq.~\eqref{eq:self}, into the computation.

To compute $\tilde \Sigma_2(0)$, we evaluate the self-energy
in Fig.~\ref{fig:b2loop} and write
\bes
-i \Sigma(p) &= g_s^2 \Cf
\int\frac{d^d k}{(2\pi)^d}
\frac{ \gamma^{\alpha}
\big(\hat p + \hat k\big) \gamma^{\beta}}{k^2\;
(k+p)^2}\times
\lp g^{\alpha\beta}   - \frac{k^\alpha k^\beta }{k^2} \rp \Pi(k^2).
\end{split}
\label{eq:sigma}
\ee
We note that, thanks to  helicity conservation, the self-energy  $\hat \Sigma$ is proportional to $\hat p$
\be
\Sigma(p) = \hat p\tilde\Sigma(p^2),~~~~
\ee
We extract $\tilde \Sigma$ from Eq.~\eqref{eq:sigma} and use dispersion relations,
Eq.~\eqref{eq:disp}, to arrive at
\be
-i\tilde\Sigma(0) = -i \asbontwopi \frac{(3-2\ep)\Gamma(1+\ep)}{(4-2\ep)(1-\ep)} \;  \frac{\Cf}{\pi}
\int \limits_{4m_b^2}^{\infty} \frac{dq^2}{q^2}
\; {\rm Im} \Pi(q^2) (q^2)^{-\ep}.
\label{eq:sigma0disp}
\ee
Combining Eq.~\eqref{eq:sigma0disp} with Eq.~\eqref{eq:b6}, we find that
\be
\lim_{q^2 \to \infty}  \left [  \frac{(3-2\ep)\Gamma(1+\ep)}{(4-2\ep)(1-\ep)} (q^2)^{-\ep}  + \widetilde{\rm Tri}(d,q^2,s) \right ]
\sim {\cal O}(q^{-2}),
\ee
which implies that in a combination of the relevant vertex correction and the wave-function renormalization contribution  the
constant asymptotic at large $q^2$ cancels out and the $q^2$ integration becomes convergent. This allows us to take the
$d \to 4$ limit in $\Pi(q^2)$.   Following this discussion, we
write the regulated $b$-quark amplitude in Eq.~\eqref{eq:ab2reg} as
\be
i\mathcal A_2^{(b\rm,reg)} =
\asbontwopi \left[\bar u_1 \gamma^\mu v_2\right]  \frac{ \Cf }{\pi}
\int\limits_{4m_b^2}^{\infty}
\frac{dq^2}{q^2} \; {\rm Im} \Pi(q^2) \;
\lp \widetilde{\rm Tri}(d=4, q^2,s) + \frac{3}{4}\rp.
\label{eq:b9}
\ee

It follows from Eq.~\eqref{eq:b9} that we only need the imaginary
part of the gluon self-energy in four dimensions.  It reads \be {\rm
  Im}\; \Pi(p^2) = \frac{2 \pi }{3}\tr \sqrt{1-\frac{4m_b^2}{p^2}}\lp
1+\frac{2m_b^2}{p^2}\rp \theta(p^2 - 4m_b^2).
\label{eq:disp4}
\ee
Inserting Eq.~\eqref{eq:disp4} into Eq.~\eqref{eq:b9}   and integrating over $q^2$, we obtain the  final result
\begin{align}
\mathcal A_2^{(b\rm,reg)} ={}&
\asontwopi^2 \Cf\tr \mathcal A_0 \, \times
\nonumber \\
\Bigg\{
&
\left(-\frac{110}{9
   (1-y)^2}+\frac{110}{9 (1-y)}-\frac{265}{54}\right) \left(\ln \frac{s}{m_b^2} -i\pi\right )+
\nonumber \\
&
\left(\frac{184}{9
   (1-y)^3}-\frac{92}{3 (1-y)^2}+\frac{56}{3 (1-y)}-\frac{38}{9}\right)
   \left[\frac{1}{2} i\pi \ln (y)+\text{Li}_2(y)+\frac{\ln ^2(y)}{4}
-\frac{\pi^2}{6}\right]
\nonumber \\
+&
\left(-\frac{8}{(1-y)^4}+\frac{16}{(1-y)^3}-\frac{8}{(1-y)^2}+
\frac{4}{3}\right)
\bigg[\frac{1}{4} i\pi \ln^2(y)+\text{Li}_3(y)+\frac{\ln ^3(y)}{12}
\nonumber \\
&
-\frac{1}{6} \pi ^2 \ln (y)-\zeta_3\bigg]
+\frac{238}{9 (1-y)^2}-\frac{238}{9 (1-y)}+\frac{3355}{324}\Bigg\},
\label{eq:b13}
\end{align}
where we have introduced two variables
\be
y = \frac{2+z-2\sqrt{1+z}}{z},~~~~ z = \frac{4m_b^2}{s}.
\ee

In hadron collisions, it is typical that $s \gg 4m_b^2$. In this case $y \approx m_b^2/s \ll 1$. We expand Eq.~\eqref{eq:b13} in powers of $y$  and find
the leading term
\be
\begin{split}
\mathcal A_2^{(b\rm,reg)} =
\asontwopi^2 \Cf\tr \mathcal A_0 \times
\Bigg  ( & \frac{1}{9} \ln^3 y
  + \left ( \frac{19}{18} + i\frac{\pi}{3} \right ) \ln^2 y
  + \left ( \frac{265}{54} - \frac{2 \pi^2}{9} + i\frac{19}{9} \right ) \ln y
  \\
&  +  \frac{3355}{324} -\frac{4}{3} \zeta_3 - \frac{19 \pi^2}{27}  + i \frac{265 \pi}{54}
  \Bigg  ) + {\cal O}(y).
\end{split}
  \ee

To conclude, we report the result for $\Pi_1(0)$, which is
required for the gluon wave-function renormalization.
From Eq.~\eqref{eq:ipimunu}, it is straightforward to obtain
\be
\Pi(0) = \asbontwopi \frac{2 T_R}{3}
\frac{\big[\Gamma(1+\ep)e^{\ep \gamma_E}\big]}{\ep}
 m_b^{-2\ep},
~~~
\label{eq:pi0}
\ee
where $S_\ep = (4\pi)^{\ep} e^{-\ep \gamma_E}$.

\bibliographystyle{utphys}
\bibliography{WHdecmass}{}

\end{document}